\documentclass[11pt]{article}
\pdfoutput=1
\usepackage{jheppub,amsmath,amssymb}
\usepackage{enumerate}
\usepackage{bm}
\usepackage{bbm}
\usepackage{enumitem}
\usepackage[usenames,dvipsnames]{xcolor}
\usepackage{amsmath}
\usepackage{amsfonts}
\usepackage{amssymb}
\usepackage{graphicx}
\usepackage{hhline}
\usepackage{subfig}
\usepackage{hyperref}
\usepackage{color}
\usepackage{verbatim}
\usepackage{amsmath,amsthm,amssymb}
\usepackage{lscape}
\usepackage{float}
\usepackage{caption}
\usepackage{lscape}
\usepackage{breqn}
\usepackage{multicol}
\usepackage{graphics}
\usepackage{tikz,todonotes}
\usepackage[utf8]{inputenc}
\usetikzlibrary{arrows,positioning,decorations.markings,decorations.pathmorphing,calc}
\pgfdeclarelayer{edgelayer}
\pgfdeclarelayer{nodelayer}
\usetikzlibrary{decorations.pathreplacing}
\pgfsetlayers{edgelayer,nodelayer,main}
\tikzset{none/.style={draw=none}}
\tikzset{new edge style 2/.style={black}}
\tikzset{new style 0/.style={black}}
\tikzset{rednode/.style={draw=none, scale=0.3pt,fill=red,circle, draw}}
\tikzset{redline/.style={line width=0.3mm,red}}
\tikzset{greyE/.style={line width=0.1mm,gray}}
\usepackage[utf8]{inputenc}
\usetikzlibrary{arrows,positioning,decorations.markings,decorations.pathmorphing,calc}

\usepackage{enumitem}
\newlist{Results}{enumerate}{2}
\setlist[Results]{label={\color{blue}\arabic*.},leftmargin=*,ref={\color{blue}\  \arabic*}}

\definecolor{hyperref}{RGB}{026,028,087}

\newcommand{\beq}{\begin{equation}}
\newcommand{\eeq}{\end{equation}}
\newcommand{\bea}{\begin{eqnarray}}
\newcommand{\eea}{\end{eqnarray}}
\def\be{\begin{equation}}
\def\ee{\end{equation}}

\def\beq{\begin{equation}}
\def\eeq{\end{equation}}

\renewcommand{\[}{\left[}
\renewcommand{\]}{\right]}

\def\be{\begin{equation}}
\def\ee{\end{equation}}
\def\ba{\begin{eqnarray}}
\def\ea{\end{eqnarray}}

\def\nn{\nonumber}

\def\d{\mathrm{d}}
\DeclareMathOperator{\tr}{tr}

\usepackage[normalem]{ulem}

\def\ba{\begin{eqnarray}}
\def\ea{\end{eqnarray}}

\def\D{\mathcal{D}}

\def\stu{St\"uckelberg }

\def\d{\mathrm{d}}

\def\({\left(}
\def\){\right)}

\begin{document}

\title{$T \bar T$ Deformations, Massive Gravity and Non-Critical Strings}

\author[a,b]{Andrew J. Tolley}
\affiliation[a]{Theoretical Physics, Blackett Laboratory, Imperial College, London, SW7 2AZ, U.K.}
\affiliation[b]{CERCA, Department of Physics, Case Western Reserve University, 10900 Euclid Ave, Cleveland, OH 44106, USA}

\emailAdd{a.tolley@imperial.ac.uk}

\abstract{The $T \bar T$ deformation of a 2 dimensional field theory living on a curved spacetime is equivalent to coupling the undeformed field theory to 2 dimensional `ghost-free' massive gravity. We derive the equivalence classically, and using a path integral formulation of the random geometries proposal, which mirrors the holographic bulk cutoff picture.  
We emphasize the role of the massive gravity \stu fields which describe the diffeomorphism between the two metrics. For a general field theory, the dynamics of the \stu fields is non-trivial, however for a CFT it trivializes and becomes equivalent to an additional pair of target space dimensions with associated curved target space geometry and dynamical worldsheet metric. That is, the $T \bar T$ deformation of a CFT on curved spacetime is equivalent to a non-critical string theory in Polyakov form, with a non-zero $B$-field. We give a direct proof of the equivalence classically without relying on gauge fixing, and determine the explicit form for the classical Hamiltonian of the $T\bar T$ deformation of an arbitrary CFT on a curved spacetime. When the QFT action is a sum of a CFT plus an operator of fixed scaling dimension, as for example in the sine-Gordon model, the equivalence to a non-critical theory string holds with a modified target space metric and modified $B$-field.
Finally we give a stochastic path integral formulation for the general $T \bar T+J \bar T+T \bar J$ deformation of a general QFT, and show that it reproduces a recent path integral proposal in the literature.

}

\maketitle


\section{Introduction}

The recent significant interest in $T \bar T $ deformations of two dimensional field theories, and their extensions,  stems from the fact that they represent a rather unique example of an irrelevant deformation of a local field theory, that may nevertheless be UV complete. Since the deformation includes a length scale, the UV theory cannot be local in the usual sense, and the deformation may be thought of as giving rise to an effectively gravitational theory \cite{Dubovsky:2012wk,Dubovsky:2013ira}. The authors of \cite{Dubovsky:2012wk,Dubovsky:2013ira} first noted this property by defining the deformed theory through its $S$-matrix via a phase factor of the Castillejo-Dalitz-Dyson (CDD) form \cite{Castillejo:1955ed}. On the field theory side, the meaningful nature of the deformation can be traced to the fact that there is a well defined composite operator \cite{Zamolodchikov:2004ce,Smirnov:2016lqw,Cavaglia:2016oda}
\be
(T \bar T)(x) = \lim_{y \rightarrow x} \frac{1}{2} \epsilon^{\mu\nu} \epsilon_{ab} T_{\mu}{}^a(x) T_{\nu}{}^b(y)= (\det[T_{\mu}{}^a])(x) \, ,
\ee
which is the operator we use to define the deformation. The $\det T$ form, applicable to any field theory has been stressed by Cardy \cite{Cardy:2018sdv}. The connection between the effective string models of \cite{Dubovsky:2012wk,Dubovsky:2013ira} and the $T \bar T$ deformation was clarified in \cite{Caselle:2013dra}. The recent intense discussion of  $T \bar T $ deformations was initiated in \cite{Smirnov:2016lqw,Cavaglia:2016oda} where explicit examples of the $T\bar T$ deformation were given. It was already noted in \cite{Cavaglia:2016oda} that the deformation of a free theory of $N$ massless bosons is classically equivalent to the Nambu-Goto action for an $N+2$ dimensional string paralleling the earlier discussion in \cite{Dubovsky:2012wk,Caselle:2013dra}. The connection with strings has also been emphasized through light cone gauge arguments \cite{Baggio:2018gct,Frolov:2019nrr,Frolov:2019xzi,Sfondrini:2019smd}. More recently, in \cite{Callebaut:2019omt}  it has been argued that the non-critical string can be used as a starting point for the definition of the $T\bar T$ deformation at the quantum level. In the analysis of the $T \bar T$ deformations, various tricks have been uncovered that allow the determination of the full $T \bar T$ deformed Lagrangian at the classical level \cite{Conti:2018tca,Conti:2018jho,Bonelli:2018kik,Coleman:2019dvf}, for which one may attempt direct quantization \cite{Rosenhaus:2019utc}. \\

Generalizations to lower spin $J \bar T$, deformations \cite{Guica:2017lia,Guica:2019vnb,LeFloch:2019rut,Anous:2019osb,Aguilera-Damia:2019tpe} and higher spin deformations have also been constructed \cite{Conti:2019dxg,LeFloch:2019wlf}. A separate effort has begun which is to understand the $T \bar T$ deformation holographically in terms of a `cutoff bulk' where in particular it is possible to discuss entanglement entropy \cite{McGough:2016lol,Bzowski:2018pcy,Guica:2019nzm,Shyam:2017znq,Gorbenko:2018oov,Dong:2018cuv,Donnelly:2019pie,Donnelly:2018bef,Araujo:2019pae,Apolo:2019yfj,Jeong:2019ylz,Grieninger:2019zts,Geng:2019ruz,Geng:2019yxo,Geng:2019bnn}. Although we will not explicitly consider these holographic approaches, the stochastic approach of section \ref{Randomgeometry} and \ref{generalstochastic} will closely mirror them by introducing an auxiliary effective spacetime dimension that describes the flow in the deformation parameter. Explicit string constructions describing the deformations are given in \cite{Giveon:2017nie,Chakraborty:2019mdf}. $T \bar T$ deformations have also been extended to non-Lorentz invariant theories \cite{Cardy:2018jho} and proposals have been made in higher dimensions \cite{Taylor:2018xcy,Hartman:2018tkw} and lower dimensions \cite{Gross:2019ach}.\\

In this note, we consider the $T \bar T$ deformation of a two dimensional field theory in curved spacetime\footnote{The expectation value of the $T \bar T$ composite operator on curved spacetimes has been explicitly considered in \cite{Jiang:2019tcq}.}, in order to address the question of what is the appropriate gravitational picture. Our analysis will be largely classical (except in section \ref{Randomgeometry} and section \ref{generalstochastic}), and so in particular we shall intentionally avoid careful discussion of the conformal anomaly. However our perspective will be that of Effective Field Theory. If two theories are equivalent quantum mechanically when the deformation is turned off $\lambda=0$, then if they are classically equivalent for finite $\lambda$, they can always be made quantum mechanically equivalent by the additional of suitable irrelevant operators that capture the effects of the path integral measure. That is because the deformation itself is an irrelevant operator, and different quantization procedures will differ only in their treatment of these irrelevant operators, provided at least that they agree for $\lambda=0$. With this in mind, we can first find the classical formulation that appears to be simplest to quantize, and address the issues of anomalies after the fact. Provided the quantum description reproduces the desired correlators for $\lambda=0$, we may regard it as a consistent definition of the deformation. We will be led to a rather obvious conclusion about what this formulation is, at least in the case where the undeformed theory is a conformal field theory (CFT). One unsung virtue of having a definition of the deformation on curved spacetime is that it allows us to define the stress energy covariantly, rather than as the Noether current and this will prove very convenient in giving simple derivations of relations that remain slightly mysterious in the current literature. \\

Briefly, we find that
\begin{itemize}
\item{The $T \bar T$ deformation of a generic field theory is equivalent to coupling the undeformed field theory to 2D `ghost-free massive gravity' \cite{deRham:2010kj}.}
\item{2D `ghost-free massive gravity' emerges naturally as the solution of a stochastic path integral for random geometries which defines the deformation at the quantum level, at least in flat space.}
\item{When the field theory is conformal, we can directly prove, by means of introducing a conformal \stu (compensator) for the deformation, the equivalence of 2D `ghost-free massive gravity' to a (in general) non-critical string theory in Polyakov form with a non-zero $B$-field.}
\item{Equivalently, starting with a non-critical string theory with curved target space metric, we may reinterpret it as a $T \bar T$ deformation of a CFT living on a generically curved spacetime.}
\item{When the undeformed theory is a CFT, we can exactly determine the classical Hamiltonian which takes the classic square root solution of the Burgers equation form.}
\item{For special classes of non-conformal field theories it is also possible to give an interpretation in terms of a non-critical string with a modified $B$-field.}
\item{We extend the stochastic path integral derivation to the general case of a $T \bar T+J \bar T+T \bar J$ deformation, and find that it reproduce a recent proposal in the literature, generalized to curved spacetime.}
\end{itemize}

Overall, these results are not entirely surprising since the connection of $T \bar T$ and Polyakov strings has been well noted for flat spacetimes \cite{Dubovsky:2012wk,Cavaglia:2016oda,Callebaut:2019omt}, in particular it is utilized in lightcone gauge in \cite{Baggio:2018gct,Frolov:2019nrr,Frolov:2019xzi,Sfondrini:2019smd}, and was implicit in earlier work \cite{Dubovsky:2012wk,Caselle:2013dra}. We will however be able to derive this directly, albeit classically, without any need to perform gauge choices, and for a curved spacetime.
Note that there is no contradiction with the observation that the gravitational descriptions of $T\bar T$ in flat spacetime are given by Jackiw-Teitelboim (topological) gravity \cite{Teitelboim:1983ux,Jackiw:1984je} (emphasized in \cite{Dubovsky:2017cnj,Dubovsky:2018bmo}) or random geometries \cite{Cardy:2018sdv,Cardy:2019qao}. Flat space Jackiw-Teitelboim (topological) gravity is equivalent to ghost-free massive gravity for a flat reference metric, and we will see in section \ref{Randomgeometry} that the massive gravity action \eqref{actionTT} can be derived directly from the path integral formulation of the  random geometry framework.
One of the striking consequences of this formalism is that we will be able to give a closed form expression for the $T \bar T$ deformation of any CFT (see section \ref{CFTHamiltonian}) on a curved spacetime. The final result for the Hamiltonian has the classic square root structure known from the solution of the Burgers equation that correctly reproduces the $T \bar T$ deformed energies of a CFT on a cylinder. Intriguingly this form is at least classically far more general. \\

We begin in section \ref{TTbartoPolyakov} with a classical derivation of the solution to the $T \bar T$ deformation, namely two dimensional ghost-free massive gravity. We then go on to to discuss some well known aspects of massive gravity, in particular the role of the \stu fields that arise in section \ref{Stuck}, the metric formulation \ref{sec:metric} and the Hamiltonian and the role of the constraint that removes the Boulware-Deser ghost \ref{HamiltonianQFT}. In the present context, it is the solution of this constraint that determines the precise form of the $T \bar T$ deformed Hamiltonian. 
In section \ref{Randomgeometry} we show that the massive gravity description emerges naturally from a path integral solution of the quantum $T \bar T$ deformation. In section \ref{CFT} we give a (classical) proof that two dimensional massive gravity coupled to a conformal field theory is equivalent to a (in general) non-critical string. In section \ref{PolyakovTTbar} we reverse the logic and show that starting with the Polyakov action for a non-critical string, it is possible to interpret it as $T \bar T$ deformation of a CFT or slightly more generally a special class of non-conformal field theories \ref{specialclass}. Finally in section \ref{generalstochastic} we give a stochastic path integral derivation of the path integral appropriate to describing the general $T \bar T+J \bar T+T \bar J$ deformation.

\section{From $T\bar T$ to Massive Gravity to Polyakov}\label{TTbartoPolyakov}

In this section our goal is to determine the classical $T\bar T$ deformation of a two dimensional field theory on a curved spacetime. We will give a quantum derivation at least valid for flat reference metrics, which may be generalizable to curved spacetimes in section \ref{Randomgeometry}.
At the classical level, the $T\bar T$ deformation of a two dimensional field theory living on a curved manifold with metric $\gamma_{\mu\nu}= f_{\mu}^a f_{\nu}^b \eta_{ab}$, and zweibein $f_{\mu}^a$ is defined by the differential relation\footnote{We use Lorentzian conventions. For convenience we work with $\epsilon^{01}=+1$ and $\epsilon_{01}=+1$ both in the Euclidean and Lorentzian so that in the Lorentzian $\epsilon^{\mu\nu} = - \eta^{\mu \alpha} \eta^{\nu \alpha} \epsilon_{\alpha \beta}$. This removes an unnecessary minus sign.}
\be\label{TTdef}
\frac{\d S_{\lambda}[\varphi,f] }{\d \lambda} =- \int \d^2 x  \,  \frac{1}{2} \epsilon^{\mu\nu} \epsilon_{ab} T_{\mu}{}^a T_{\nu}{}^b =-\int \d^2 x \det[T_{\mu}{}^a] \, ,
\ee
where $\varphi$ denotes the unspecified matter fields. Here the stress energy is that of the deformed theory, with deformation parameter $\lambda$, defined in mixed Lorentz and diffeomorphism indices by
\be
\det(f) \, T^{\mu}{}_a(x) = \frac{\delta S_{\lambda}[\varphi,f]}{\delta f_{\mu}{}^a(x)} \, .
\ee
The above differential equation appears to be complicated to solve in general, and original attempts made use of well chosen ansatz or other tricks \cite{Cavaglia:2016oda,Bonelli:2018kik,Coleman:2019dvf}. However, it is possible to give a practical general solution by means of introducing an auxiliary zweibein $e_{\mu}^a$ (and associated metric  $g_{\mu\nu}= e_{\mu}^a e_{\nu}^b \eta_{ab}$) and considering the action
\be\label{actionTTa}
S_{T\bar T}[\varphi,f,e] = \int \d^2 x  \,  \frac{1}{2 \lambda} \epsilon^{\mu\nu} \epsilon_{ab} (e_{\mu}^a- f_{\mu}^a) (e_{\nu}^b- f_{\nu}^b) + S_{0}[\varphi,e]
\ee
where $S_{0}[\varphi,e]$ is the undeformed action living on the spacetime defined by the zweibein $e_{\mu}^a$. We will often refer to this as the seed theory. The stress energy defined from $S_{T\bar T}$ by varying with respect to $f$ gives
\be\label{efrelation}
\det(f) \, T^{\mu}{}_a[\varphi,f,e]  = -\frac{1}{\lambda} \epsilon^{\mu\nu} \epsilon_{ab} (e_{\nu}^b- f_{\nu}^b) \, ,
\ee
which is better written as
\be\label{efrelation2}
e_{\mu}{}^a= f_{\mu}{}^a-\lambda \det f \epsilon_{\mu\nu} \epsilon^{ab}  T^{\nu}{}_{b}[\varphi,f,e] \, .
\ee
We stress this is not an equation of motion, but a definition of the stress energy tensor. The actual equation of motion is the deceptively similar equation
\be \label{eom1}
\frac{1}{\lambda} \epsilon^{\mu\nu} \epsilon_{ab} (e^b_{\nu}- f^{b}_{\nu})+ \frac{\delta S_0[\varphi,e]}{\delta e_{\mu}^a} =0\, .
\ee
In the absence of curvature couplings, this equation is an algebraic equation for the zweibein $e$ which may be solved relatively straightforwardly. We denote by $e_*$ the associated on-shell value of $e_{\mu}{}^a$ which is now $\lambda$ dependent by virtue of equation \eqref{eom1}. In turn from this equation we find a simple relationship between the stress energy of the undeformed theory on the curved geometry $e$ and that of the $T \bar T$ deformed theory, namely
\be\label{TTrelation}
\det f T^{\mu}{}_a =  \frac{\delta S_0[\varphi]}{\delta e_{\mu}^a} \Big|_{e=e_*} = \det e_* {T_{0}}^{\mu}{}_a(\varphi,e_*)\, .
\ee
Remembering that the spacetime indices on $T^{\mu}{}_a$ are raised and lowered using $f/\gamma$ and those in ${T_{0}}^{\mu}{}_a$ using $e/g$ then by taking the determinant of both sides we find
\be
\det  T_{\mu}{}^a = \det  {T_{0}}_{\mu}{}^a \, .
\ee
To see that the action \eqref{actionTTa} is correctly defining the $T\bar T$ deformation we note that
\be
\frac{\d S_{T\bar T}[\varphi,f,e_*] }{\d \lambda} =- \int \d^2 x  \,  \frac{1}{2 \lambda^2} \epsilon^{\mu\nu} \epsilon_{ab} (e_{\mu}^a- f_{\mu}^a) (e_{\nu}^b- f_{\nu}^b) + \int \d^2 x \frac{\d e_*}{\d \lambda} \frac{\delta S_{T\bar T}[\varphi,f,e_*]}{\delta e} \, .
\ee
The last term vanishes, by virtue of the on-shell condition, and so we have on-shell
\ba
\frac{\d S_{T\bar T}[\varphi,f,e_*] }{\d \lambda} &=& - \int \d^2 x  \,  \frac{1}{2 \lambda^2} \epsilon^{\mu\nu} \epsilon_{ab} (e_{*\mu}{}^a- f_{\mu}{}^a) (e_{*\nu}{}^b- f_{\nu}{}^b) \\
&=&- \int \d^2 x  \,  \frac{1}{2} \epsilon^{\mu\nu} \epsilon_{ab} T_{\mu}{}^a[\varphi,f,e_*]  T_{\nu}{}^b[\varphi,f,e_*]  \, .
\ea
This is precisely the defining relation \eqref{TTdef} given the identification
\be
S_{\lambda}[\varphi,f]= S_{T\bar T}[\varphi,f,e_*] \, ,
\ee
together with the on-shell equivalence of the two stress energies $T_{\mu}{}^{a}  = T_{\mu}^a[\varphi,f,e_*] $ from
\be
 \frac{\delta S_{\lambda}[\varphi,f]}{\delta f_{\mu}{}^a(x)} = \frac{\delta S_{T\bar T}[\varphi,f,e_*]}{\delta f_{\mu}{}^a(x)} +  \int \d^2 x' \frac{\delta e_{*,\nu}{}^b(x')}{\delta f_{\mu}{}^a(x)} \frac{\delta S_{T\bar T}[\varphi,f,e_*]}{\delta e_{\nu}{}^b(x')} \, .
\ee
Hence classically at least the $T \bar T$ deformation of an undeformed action $S_{0}[\varphi,f]$ is completely described by the action\footnote{Interestingly this Lagrangian appeared in \cite{Freidel:2008sh} as a tool to determine the solution of the Wheeler-de Witt equation in the context of the $AdS^3/CFT^2$ correspondence. The connection with $T \bar T$ was noted in \cite{McGough:2016lol}.}
\be\label{actionTT}
S_{T\bar T}[\varphi,f,e] = \int \d^2 x  \,   \frac{m^2}{2 } \epsilon^{\mu\nu} \epsilon_{ab} (e_{\mu}^a- f_{\mu}^a) (e_{\nu}^b- f_{\nu}^b) + S_{0}[\varphi,e] \, .
\ee
where we identify $m^2=1/\lambda$. Remarkably the resulting gravitational theory, where $e$ is viewed as the dynamical metric, is well known, it is simply `ghost-free' massive gravity\footnote{Sometimes referred to as the de Rham-Gabadadze-Tolley (dRGT) model. } in two dimensions \cite{deRham:2010kj} (see \cite{deRham:2014zqa} for an extensive review). In the parlance of massive gravity, $e$ is the dynamical zweibein and $f$ is the reference zweibein. This connection may be made more explicit by putting the above action in a pure metric form, as we do below in section \ref{sec:metric}, as was originally done in \cite{deRham:2010kj}. The vielbein form of ghost-free massive gravity which corresponds to \eqref{actionTT} was given in \cite{Hinterbichler:2012cn}.

\subsection{Classical Trace Flow equation}

If the undeformed theory is a CFT, then it is possible to derive a simple relation that describes the flow of the stress energy tensor. We denote the trace of the stress energy tensor by
\be
\Theta = f_{\mu}{}^a T^{\mu}{}_a = \frac{1}{\det f} f_{\mu}{}^a \frac{\delta S_{\lambda}[\varphi,f]}{\delta f_{\mu}{}^a(x)} \, .
\ee
The vanishing of the trace of the stress energy for the undeformed theory $ \frac{\delta S_0[\varphi]}{\delta e_{\mu}^a} e_{\mu}^a=0$ together with the equation of motion for $e$ \eqref{eom1} implies that
\be
\epsilon^{\mu\nu} \epsilon_{ab} e^{a}_{\mu}(e^b_{\nu}- f^{b}_{\nu}) =0 \, .
\ee
Using \eqref{efrelation}, this is the statement that
\be
e_{\mu}{}^a T^{\mu}{}_a = (f_{\mu}{}^a-\lambda \det f \epsilon_{\mu\nu} \epsilon^{ab}  T^{\nu}{}_{b} ) T^{\mu}{}_a  =0 \, ,
\ee
in other words the trace in the deformed theory is  
\be\label{Trace}
\Theta =  2 \lambda \det f \det(  T^{\mu}_{a} )=2 \lambda \det(  T^{\mu}_{\nu} )= - \lambda \( T_{\mu \nu} T^{\mu\nu}- \Theta^2 \) \, .
\ee
Given this the deformation equation is simply the response of the action to a change of scale reflecting the fact that for a classical CFT, the only scale that arises is the deformation parameter itself
\be\label{TTdef2}
\frac{\d S_{\lambda}[\varphi,f] }{\d \lambda} =-\int \d^2 x  \,  \frac{1}{2 \lambda}  \det f \, \Theta(x) \, .
\ee
Eq.~\eqref{Trace} is of course a classical relation, and we expect it to be modified at the quantum level by at least the presence of the conformal anomaly to something of the form
\be\label{Trace2}
\Theta = - \frac{c}{24 \pi} R[f] - \lambda \( T_{\mu \nu} T^{\mu\nu}- \Theta^2 \) \,  + \dots \, ,
\ee
where $\dots$ may account for higher derivative corrections. The precise modification will be determined by the precise quantization prescription. Specific proposals have been given guided by holography \cite{McGough:2016lol,Hartman:2018tkw,Guica:2019nzm}, Our perspective will rather be to first find the best classical description and worry about the anomaly afterwards.

\subsection{\stu Fields}\label{Stuck}

Although there are no propagating massive gravitons in two dimensions, the sense in which this is massive is that the fixed zweibein $f$ spontaneously breaks the symmetries of the gravitational theory. Two vielbeins $e$ and $f$ both transform in principle under separate copies of diffeomorphisms and local Lorentz transformations. When they are coupled together in the action \eqref{actionTT}, these symmetries are spontaneously broken down to a single diagonal copy 
\be
(Diff[M] \times{\rm Lorentz}) \times (Diff[M] \times{\rm Lorentz}) \rightarrow  Diff[M]_{\rm diag} \times {\rm Lorentz}_{\rm diag} \, .
\ee
In other words it is invariant under local Lorentz transformations and diffeomorphisms that act identically on $e$ and $f$
\ba
&& e_{\mu}^a(x) \d x^{\mu} =\lambda^a{}_b(x') {e'}_{\mu}^b(x') \d {x'}^{\mu} \, ,  \\
&& f_{\mu}^a(x) \d x^{\mu} =\lambda^a{}_b(x') {f'}_{\mu}^b(x') \d {x'}^{\mu}  \, ,
\ea
where $\lambda^a{}_b(x')$ denotes a local Lorentz transformation $\lambda^T \eta \lambda = \eta$. This would be the symmetry were both $e$ and $f$ regarded as dynamical. However, when the reference zweibein is in turn taken to be non-dynamical, these remaining local symmetries are broken down to the global isometries of $f$. 
\be
 Diff[M]_{\rm diag} \times {\rm Lorentz}_{\rm diag} \rightarrow {\rm Isom}(f) \, ,
\ee
in other words the spacetime symmetries of the undeformed action. This is the same breaking pattern that occurs in massive gravity in any dimension \cite{Ondo:2013wka,Fasiello:2013woa,Gabadadze:2013ria}. As is standard, it is convenient to describe the broken state by reintroducing \stu fields for the broken symmetries \cite{ArkaniHamed:2002sp}. The \stu fields are the fields that become the Goldstone modes in the global limit. 

In the present case, to recover the local $Diff[M]_{\rm diag} \times {\rm Lorentz}_{\rm diag} $ symmetry we introduce local Lorentz $\Lambda^a{}_b(x) = e^{\eta^{ac}(x)\omega_{cb}}$ and diffeomorphism $\Phi^A(x)$ \stu fields (we follow the conventions of \cite{Ondo:2013wka,Fasiello:2013woa,Gabadadze:2013ria}). Since we are in two dimensions there is only one Lorentz \stu field $\omega_{ab}= \epsilon_{ab}\omega$, and we have two diffeomorphism \stu fields $\Phi^A$.  The reference zweibein takes the form
\be
f_{\mu}^a(x) = \Lambda^a{}_b \partial_{\mu} \Phi^A F_{A}{}^b(\Phi)
\ee
such that
\be
\gamma_{\mu\nu} \d x^{\mu} \d x^{\nu} = \eta_{ab}F_{A}{}^a(\Phi)F_{B}{}^b(\Phi) \d \Phi^A \d \Phi^B= \hat \gamma_{AB}(\Phi)  \d \Phi^A \d \Phi^B \,  .
\ee
For example, in the simplest case in which the reference metric is Minkowski we have $F_A^a = \delta_A^a$ and so
\be
f_{\mu}^a = \Lambda^a{}_b\partial_{\mu} \Phi^b \, .
\ee
If we work in unitary gauge $\Lambda^a{}_b=\delta^a_b$, i.e. $\omega=0$, for the Lorentz \stu fields, but not for the diffeomorphism \stu fields, then we get the form of the action determined in \cite{Dubovsky:2017cnj,Dubovsky:2018bmo} which is equivalent to the (flat space) Jackiw-Teitelboim gravity \cite{Teitelboim:1983ux,Jackiw:1984je}. More generally it is clear however that it is better to interpret this action as that for massive gravity in order to describe field theories on arbitrary spacetimes. \\

\subsection{Metric Formulation} \label{sec:metric} 

The full \stu form of the action \eqref{actionTT} is
\be
S_{T\bar T}[\varphi,f,e] = \int \d^2 x  \,   \frac{m^2}{2 } \epsilon^{\mu\nu} \epsilon_{ab} (e_{\mu}^a-\Lambda^a{}_c \partial_{\mu} \Phi^A F_{A}{}^c(\Phi)) (e_{\nu}^b- \Lambda^b{}_d \partial_{\nu} \Phi^B F_{B}{}^d(\Phi)) + S_{0}[\varphi,e] \, .
\ee
By virtue of the \stu fields this is now manifestly diffeomorphism and local Lorentz invariant. Crucially the matter fields $\varphi$ do not directly couple to the \stu fields, and furthermore the Lorentz \stu fields arise as auxiliary variables which may be integrated out. To do this we note that the equation of motion for the Lorentz \stu field is the so-called `symmetric vielbein' condition \cite{Hinterbichler:2012cn,Ondo:2013wka}
\be\label{symmetriczweibein}
\frac{\delta S_{T\bar T}}{\delta \omega_{ab}}=0 \quad \rightarrow \quad e_{\mu}^a f_{\nu}^b \eta_{ab} = e_{\nu}^a f_{\mu}^b \eta_{ab} \, ,
\ee
or in shorthand
\be
(e^T \eta f)  =(e^T \eta f)^T = f^T \eta e \, . 
\ee
In two dimensions, this is only one independent equation, for one independent Lorentz \stu $\omega$.
Explicitly isolating the Lorentz \stu fields then we have more concisely $f= \Lambda \hat F$, i.e. $f^a_{\mu} = \Lambda^a _b \hat F_{\mu}^b$ and $\hat F_{\mu}^a = F^a_{A} \partial_{\mu} \Phi^A$, so that the symmetric vielbein condition is
\be
e^T \eta \Lambda \hat F= \hat F^T \Lambda^T \eta e \, . 
\ee
The solution of this equation is well known, and is obtained from constructing the combination (using $\Lambda^T \eta \Lambda=\eta$)
\be
e^T \eta \(  \Lambda \hat F e^{-1} \)^2 = \hat F^T \eta \hat F e^{-1} \, ,
\ee
and then square rooting and using a similarity transformation\footnote{Note that the square root is unambigously defined by diagonalizing $g^{-1}f$ with a similarity transformation, which is always possible given the symmetric nature of tensors, and then taking the square root of each of its eigenvalues with positive sign. }
\be\label{squareroot}
\Lambda = \(  \sqrt{\eta^{-1}  {e^T}^{-1} \hat F^T \eta \hat F e^{-1} } \)  e \hat F^{-1} = e \sqrt{g^{-1}  \gamma} \hat F^{-1} \, ,
\ee
where we have written the square roots in terms of the metrics $g = e^T \eta e$ and $\gamma = \hat F^T \eta \hat F$. 
Crucially then
\be
e^{-1} f = e^{-1} \Lambda \hat F = \sqrt{g^{-1}  \gamma} \, .
\ee
This is the origin of the square root structure characteristic of ghost-free massive gravity \cite{deRham:2010kj,Hinterbichler:2012cn}.
Now 
\be
\det[e_\mu^a-f_{\mu}^a] = \det(e) \det[1-e^{-1} f] = \sqrt{-\det g} \det[1-\sqrt{g^{-1}  \gamma}] \, .
\ee
and on substituting back in the action, we obtain 
\be\label{metric}
S_{T\bar T}[\varphi,\gamma,e] = \int \d^2 x  \sqrt{-\det g} \[  - \frac{1}{2} m^2 ({\rm Tr}[K^2]- {\rm Tr}[K]^2)\]+ S_{0}[\varphi,e] \, ,
\ee
where 
\be
K^{\mu}{}_{\nu} =  \delta^{\mu}_{\nu} -\sqrt{g^{\mu \alpha} \gamma_{\alpha \nu} } =\delta^{\mu}_{\nu} -\sqrt{g^{\mu \alpha} \hat \gamma_{AB}(\Phi) \partial_{\mu} \Phi^A \partial_{\nu} \Phi^B} \, .
\ee
This is precisely the metric form of massive gravity that was explicitly considered in \cite{deRham:2010kj} for a Minkowski reference metric.

\subsection{Explicit Form for Mass Term} 

The unwieldy nature of the matrix square roots prompts us to put the above action in a more explicit form. Fortunately in two dimensions this is easy to do. We note that
\ba
\det K= \det\(I-\sqrt{g^{-1}  \gamma}\) &=& 1 - \tr[\sqrt{g^{-1}  \gamma}] + \det\(\sqrt{g^{-1} \gamma}\)  \, , \\
&=&1 - \tr[\sqrt{g^{-1}  \gamma}] + \sqrt{(\det g)^{-1} \det( \gamma)} \, .
\ea
Then using 
\be
\det\(\sqrt{g^{-1} \gamma}\)= \sqrt{(\det g)^{-1} \det( \gamma)} = \frac{1}{2} \( \tr[\sqrt{g^{-1} \gamma}] \)^2-\frac{1}{2} \tr[g^{-1}  \gamma] \, ,
\ee
and rearranging
\be
\tr[\sqrt{g^{-1} \gamma}]  =\sqrt{\tr[g^{-1} \gamma]  + 2 \sqrt{(\det g)^{-1} \det( \gamma)} } \, .
\ee
Hence the action is given more explicitly as
\be
S_{T\bar T}[\varphi,\gamma,e] = \int \d^2 x  \,{\cal L}_{\rm Stuck}[\Phi,g]   + S_{0}[\varphi,e] \, ,
\ee
where \stu Lagrangian has a highly unusual non-linear sigma model form:
\be\label{massterm}
{\cal L}_{\rm Stuck}[\Phi,g,\gamma] =m^2  \(  \sqrt{-\det g} +  \sqrt{-\det \gamma} -\sqrt{ -\det g \tr[g^{-1} \gamma]  + 2 \sqrt{-\det g}\sqrt{-\det( \gamma)} } \) \, .
\ee
This would be relatively straightforward were it not for the third square root structure. For massive gravity in higher dimensions, this \stu Lagrangian becomes increasingly more complicated, and has been considered explicitly in for example \cite{deRham:2011rn,Alberte:2013sma,deRham:2015ijs,deRham:2016plk,deRham:2012kf,deRham:2018svs}. In higher dimensions, the \stu fields describe the extra helicity degrees of freedom in a massive graviton, namely the helicity-one and helicity-zero modes \cite{deRham:2011qq,deRham:2011rn}. In two dimensions, even a massive graviton has no propagating degrees of freedom, and so despite appearances ${\cal L}_{\rm Stuck}[\Phi,g,\gamma] $ does not describe two scalar degrees of freedom \cite{Alberte:2013sma}. The special square root structure preserves a symmetry (for flat reference metric) which kills off the would-be dynamics of the two scalars $\Phi^a$  \cite{Alberte:2013sma} and more generally a pair of second class constraints. Establishing this symmetry or constraints is notoriously difficult (see \cite{deRham:2011rn,Alberte:2013sma}), however in unitary gauge in two dimensions it becomes more straightforward as we see in section \ref{HamiltonianQFT}. 

\subsection{$T \bar T$ Hamiltonian for a generic QFT}\label{HamiltonianQFT}

The Hamiltonian for the $T \bar T$ deformation, i.e. that for two dimensional massive gravity, for a Minkowski reference metric was already worked out in \cite{deRham:2010kj} as a toy example of the `ghost-free' massive gravity structure (see also \cite{Alberte:2013sma}). We repeat the argument here generalized to curved reference metric (in higher dimensions this Hamiltonian analysis was considered in \cite{Hassan:2011tf}). We may always put the two metrics in ADM form
\be
g_{\mu\nu} \d x^{\mu} \d x^{\nu}  = - N^2(x,t) \d t^2 +  A^2(x,t) ( \d x+ N^x(x,t) \d t)^2 \, ,
\ee
and
\be
\gamma_{\mu\nu} \d x^{\mu} \d x^{\nu} = - M^2(x,t)  \d t^2 + B^2(x,t)  ( \d x+ M^x(x,t)  \d t)^2 \, ,
\ee
where in this subsection we work in unitary gauge $\Phi^a = x^a$.

The covariant action of the undeformed theory can always be written in the canonical form
\be\label{canonicalaction}
 S_{0}[\varphi,e] = \int \d^2 x \left[  \sum_I \Pi_I \dot \varphi_I - N {\cal H} - N^x {\cal H}_x- \sum_{a} \lambda_a C_a\right] \, ,
\ee
where ${\cal H}$ is the Hamiltonian constraint and ${\cal H}_x$ is the momentum constraint, which are functions of the phase space variables $\varphi_I$, $\Pi_I$ and their spatial derivatives. $\Pi_I$ are the momenta conjugate to the set of matter fields $\varphi_I$, about which we have made no assumptions. In particular this expression would be equally valid for fermionic fields since its structure is fixed by diffeomorphism invariance.  The $\lambda_a$ are Lagrange multipliers for any other possible constraints $C_a=0$ in the system. More generally, when there are for example curvature couplings, for instance for Einstein-dilaton gravity, we may need to introduce a momentum conjugate to $A$, i.e. a term $ \int \d^2 x  \, \Pi_A \dot A$, however this does not arise in the pure classical ghost-free massive gravity where the Einstein-Hilbert term is purely topological.

In order to evaluate the mass term \eqref{massterm} we need
\be
\sqrt{-\det g} =NA \, ,  \quad \sqrt{-\det \gamma} = MB \, , \quad \tr[g^{-1} \gamma] = \frac{B^2 N^2+A^2 (M^2 - B^2 (M^x-N^x)^2}{A^2 N^2} \, .
\ee
Hence the mass term evaluated in unitary gauge is
\ba
{\cal L}_{\rm Stuck}[x^a,g,\gamma] &=& m^2 \(NA + MB - NA \sqrt{ 2 \frac{MB}{NA}+\frac{B^2 N^2+A^2 (M^2 - B^2 (M^x-N^x)^2}{A^2 N^2}}\) \nn \\
&=& m^2 \(NA + MB -\sqrt{ (BN +AM)^2-A^2 B^2 (M^x-N^x)^2}\) \, .
\ea
As noted in \cite{deRham:2010kj} the structure of the canonical action is considerably simplified with the change of variables $N^x = M^x + (BN+AM) n^x/(AB)$ so that it takes the form
\ba
S_{T \bar T} &=& \int \d^2 x \left[  \sum_I \Pi_I \dot \varphi_I - N {\cal H} - M^x {\cal H}_x- \sum_{a} \lambda_a C_a + m^2 (NA+MB) \right. \nn  \\
&& \left. - (NB+MA) \(\frac{1}{AB} n^x {\cal H}_x+ m^2 \sqrt{1-(n^x)^2} \) \] \, .
\ea
The special magic of `ghost-free' massive gravity \cite{deRham:2010kj} is that it is now possible to integrate out the non-dynamical shift $n^x$ and leave behind an action which is linear $N$, so that $N$ acts as a Lagrange multiplier for the constraint that removes the Boulware-Deser (BD) ghost \cite{Boulware:1973my}. Explicitly we have
\be
n^x = \frac{\frac{{\cal H}_x}{m^2 AB}  }{\sqrt{1+ \(\frac{{\cal H}^x}{m^2 AB}  \)^2}} \, ,
\ee
and so substituting back in gives
\be
S_{T \bar T} = \int \d^2 x \left[  \sum_I \Pi_I \dot \varphi_I - N C_{BD} - M^x {\cal H}_x- \sum_{a} \lambda_a C_a   - m^2 M \(  \sqrt{A^2+ \(\frac{m^2{\cal H}^x}{B}\)^2}-B\)  \] \, ,
\ee
where the BD ghost removing constraint $C_{BD}$ becomes
\be\label{BDconstraint}
C_{BD}={\cal H} +m^2 B  \sqrt{1+ \(\frac{{\cal H}^x}{m^2 AB}  \)^2}-A m^2 =0\, .
\ee
This constraint should be viewed as an equation for $A$. Indeed, in the case where the original Lagrangian also included a $ \int \d^2 x  \, \Pi_A \dot A$ term, this constraint will remove the dynamics of $A$ and ultimately that of $\Pi_A$. We denote the solution of \eqref{BDconstraint} by $A_*$. For a general field theory we need to specify the Lagrangian in order to determine $A_*$. Nevertheless, assuming this is known we can give an expression for the Hamiltonian evaluated on the constraint surface $C_a=C_{BD}=0$ (assuming $ \int \d^2 x  \, \Pi_A \dot A$ term is absent) namely
\be\label{Ham1}
H = \int \d x \[ m^2 M \( \sqrt{A_*^2+ \(\frac{{\cal H}^x(A_*)}{m^2 B}  \)^2} -B\)+ M^x {\cal H}_x(A_*) \] \, ,
\ee
where ${\cal H}_x(A_*)$ is understood to be the momentum constraint evaluated on the solution of the BD constraint. 

To confirm that we have the correct solution we note that for a massless minimally coupled scalar $\varphi$ for which $ S_{0}[\varphi,e]= \int \d^2 x \( -\frac{1}{2} \sqrt{-\det g} g^{\mu\nu}\partial_{\mu} \varphi \partial_{\nu} \varphi \) $ the Hamiltonian and momentum constraints are
\be
{\cal H} = \frac{1}{2 A} \Pi^2 + \frac{1}{2A} ( \partial_x \varphi)^2 \, , \quad {\cal H}^x = \Pi \partial_x \varphi \, ,
\ee
and on solving the BD constraint we recover the correct two dimensional Nambu-Goto Hamiltonian on curved spacetime which is indeed the correct $T \bar T $ deformation.
\be
H = \int \d x \[ \frac{m^2 M B}{2} \( -1 + \sqrt{\( 1+ \frac{2 \Pi^2}{m^2 B^2}\) \(1+ \frac{2 ( \partial_x \varphi)^2}{m^2 B^2} \)}\)+ M^x \Pi \partial_x \varphi \] \, .
\ee
For a general field theory we cannot be more explicit about the form of the Hamiltonian than \eqref{Ham1} without specifying the Lagrangian. However the situation is very different when the seed theory is a CFT. This is essentially because for a CFT, the $A$ dependence of the Hamiltonian is fixed by conformal symmetry and so it is trivial to solve the BD constraint. We shall give the $T \bar T$ Hamiltonian for a general CFT in section \ref{CFTHamiltonian} arriving at it by a slightly different but equivalent method.

\subsection{Random Geometries and a Stochastic Path Integral}\label{Randomgeometry}

If we define perturbations of the metric as
\be
g_{\mu\nu} = \gamma_{\mu\nu} + 2 h_{\mu\nu}
\ee
then at quadratic order the above action takes the standard Fierz-Pauli form 
\be
S_{T\bar T}^{(2)} = \int \d^2 x  \sqrt{-\det \gamma} \[  - \frac{1}{2} m^2 \( h_{\mu\nu}h^{\mu\nu}-h^2 \)+ h_{\mu\nu} T_0^{\mu\nu} + \dots \] \, .
\ee
where indices are now raised and lowered with respect to the reference metric. The only difference between this and the higher dimensional generalization is the absence of an explicit kinetic term for the massive spin-2 state $h_{\mu\nu}$ reflected in the fact that even massive gravitons in two dimensions have no propagating degrees of freedom. Within the path integral over metrics, this gives the Hubbard-Stratonovich transformation discussed in \cite{Cardy:2018sdv} which reproduces the leading order $T\bar T$ deformation. We may thus regard the metric formulation \eqref{metric} or the equivalent zweibein formulation \eqref{actionTT} as the correct nonlinear generalization of the Hubbard-Stratonovich transformation, generalized to curved spacetimes. \\

Using the result of section \ref{Fielddiffs}, when the reference metric $f$ is Minkowski, then $e$ is also forced to be Minkowski in the same local Lorentz frame, but a different diffeomorphism frame. Thus on-shell we may gauge fix $e^a_{\mu} = \delta^a_{\mu}$ and $f^a_{\mu} = \partial_{\mu} \Phi^a$. Then denoting $\Phi^a = x^a + \pi^a$, the \stu action evaluated on-shell is
\ba
S_{\text{\stu}} &=& \int \d^2 x \frac{m^2}{2}\epsilon^{\mu \nu} \epsilon_{ab} (e_{\mu}^a-f_{\mu}^a) (e_{\nu}^b-f_{\nu}^b)  \\
&=&\int \d^2 x \frac{m^2}{2}\epsilon^{\mu\nu} \epsilon_{ab}  \partial_{\mu} \pi^a \partial_{\nu} \pi^b \\
&=&  \int_{\partial M} \d x_{\mu} \frac{m^2}{2}\epsilon^{\mu\nu} \epsilon_{ab}   \pi^a \partial_{\nu} \pi^b \, , \
\ea
which is a pure boundary term. Cardy \cite{Cardy:2018sdv} argues that it is this total derivative feature that is at the root of the solvability of the $T \bar T$ deformation. This is also why in the flat case, this theory is often referred to as topological gravity. This simple total derivative property does not appear to extend to curved reference metrics, and so it is not entirely clear these arguments are useful more generally. This appears to correlate with the more complicated properties of the expectation value of the composite $T \bar T$ operator on curved spacetime \cite{Jiang:2019tcq}. \\

The random geometries framework of  \cite{Cardy:2018sdv}, appropriately formulated in terms of zweibeins, can be used to give an explicit path integral derivation of \eqref{actionTT}. At the quantum level, the $T \bar T$ deformation can be defined for the path integral (partition function in the Euclidean) via the functional differential equation\footnote{Note that since in two dimensions $\det( T_{\mu}^a)= \det\( (\det f )T^{\mu}_{a} \)$, no factors of $\det f$ are needed in \eqref{functionalequation} which is crucial to the simplicity of its solution. Furthermore the superspace Laplacian has a trivial measure as explained in Appendix~\ref{app:measure}.}
\be\label{functionalequation}
i \frac{\d Z_{\lambda}[f,J]}{\d \lambda} = \int \d^2 x  \lim_{y \rightarrow x}  \,  \frac{1}{2}  \epsilon_{\mu\nu} \epsilon^{ab} (- i)^2\frac{\delta^2 Z_{\lambda}[f,J]}{\delta f_{\nu}^b(x)\delta f_{\mu}^a(y) } \, ,
\ee
here $J$ denotes sources for the matter fields $\varphi$. It is possible that this equation should be supplemented by additional curvature terms, so that the limit $y \rightarrow x$ is well defined, i.e. contact terms, but for now we assume this is the correct starting point.
This takes the form of a stochastic differential equation\footnote{In the Euclidean this is a special case of the Fokker-Planck equation, with the right hand side being the diffusion term. Although we will not use it, the solutions can be put in Langevin form.}, or a field theory functional Schr\"odinger equation, depending on Euclidean versus Lorentzian perspective. The Hamiltonian in this Schr\"odinger equation is purely quadratic in momenta, only because we are working with the zweibeins as the fundamental variable, and we are in two dimensions.  \\

Once this is recognized, we see that \eqref{functionalequation} is straightforward to solve formally in terms of a path integral over fields which are functions of $(x,\lambda)$, i.e. in which $\lambda$ is effectively viewed as an additional spacetime dimension. Introducing $P^{\mu}_a $ as a momentum conjugate to $f_{\mu}^a$ in this higher dimensional Hamiltonian, then we have the well known canonical path integral as the solution
\be
 Z_{\lambda_f}[f(\lambda_f),J] = \int D f[x,\lambda] \int D P[x,\lambda] e^{i \int_{0}^{\lambda_f} \d \lambda \int \d^2 x   \left[ P^{\mu}_a \partial_{\lambda} f_{\mu}^a  -\frac{1}{2}  \epsilon_{\mu\nu} \epsilon^{ab} P^{\mu}_a  P^{\nu}_b \right] }Z_{0}[f(0),J] \, , 
\ee
where $Z_{0}[f(0),J] $ is the undeformed path integral. Since the exponent of the path integral is linear in $f_{\mu}^a(x,\lambda)$, we may integrate it out, leaving only the path integral over the initial zweibein $f_{\mu}^a(x,0)$. Being careful to maintain the boundary terms, we find
\ba\label{stochasticpath}
 Z_{\lambda_f}[f(\lambda_f),J] &=& \int D f[x,0]   \int D P[x,\lambda]  \delta\(\frac{\partial P[\lambda]}{\partial \lambda} \) e^{i \int \d^2 x   \left[  \( P^{\mu}_a(\lambda_f)  f_{\mu}^a(\lambda_f) -P^{\mu}_a(0)  f_{\mu}^a(0) \) \right]} \nn   \\
 && e^{i \int_{0}^{\lambda_f} \d \lambda \int \d^2 x   \left[  -\frac{1}{2}  \epsilon_{\mu\nu} \epsilon^{ab} P^{\mu}_a  P^{\nu}_b \right] }Z_{0}[f(0),J]  \, .
\ea
Since the path integral delta function enforces $\frac{\partial P^{\mu}_a[x,\lambda]}{\partial \lambda} =0$, the conjugate momentum path integral reduces to that over the initial momentum only which we just denote $P_{\mu}^a(x,\lambda=0)=P_{\mu}^a(x)$,
\be
 Z_{\lambda_f}[f(\lambda_f),J] = \int D f[x,0] \int D P[x]  e^{i  \int \d^2 x   \left[ P^{\mu}_a \(  f_{\mu}^a(\lambda_f) -f_{\mu}^a(0) \) \right]}  e^{i \lambda_f \int \d^2 x   \left[  -\frac{1}{2}  \epsilon_{\mu\nu} \epsilon^{ab} P^{\mu}_a  P^{\nu}_b  \right] }Z_{0}[f(0),J] \, .
\ee
To make the notation more transparent, we now denote $\lambda_f=\lambda$, $f_{\mu}^a(0)= e_{\mu}^a$ and $f_{\mu}^a(\lambda_f)= f_{\mu}^a$ so that we have
\be
 Z_{\lambda}[f,J] = \int D e[x] \int D P[x] \,  e^{i  \int \d^2 x   \left[P^{\mu}_a\(  f_{\nu}^b -e_{\nu}^b \) \right]}  e^{i \lambda \int \d^2 x   \left[  -\frac{1}{2}  \epsilon_{\mu\nu} \epsilon^{ab} P^{\mu}_a  P^{\nu}_b  \right] }Z_{0}[e,J] \, .
\ee
Performing the final Gaussian integral, and ignoring trivial measure factors, we finally have
\be\label{result}
 Z_{\lambda}[f,J] = \int D e[x] \,  e^{i  \int \d^2 x   \left[ \frac{1}{2 \lambda} \epsilon^{\mu\nu} \epsilon_{ab} \(  e_{\mu}^a -f_{\mu}^a \) \(  e_{\nu}^b -f_{\nu}^b \) \right]}  Z_{0}[e,J] \, .
\ee
This is the quantum version of the central result \eqref{actionTT}, that the $T\bar T$ deformation is equivalent, to taking the original field theory, coupling it to a dynamical spacetime $e$, with the addition of a massive gravity mass term. The validity of this result depends on the validity of \eqref{functionalequation} and whether or not we need to add additional curvature terms to deal with the coincidence limit. The arguments of \cite{Dubovsky:2017cnj,Dubovsky:2018bmo} which apply when $f$ is flat strongly suggest that in this case \eqref{result} is indeed the correct result. \\

In writing \eqref{result} we have assumed a standard linear measure on the path integral over $e$. In fact as explained in Appendix~\ref{app:measure}, this is the correct diffeomorphism invariant measure that reproduces the standard Polyakov measure when written in terms of the metric. Furthermore, this measure is consistent with the requirement that as $\lambda \rightarrow 0$ we recover the original undeformed theory. To see this we change variables and write $e=f+\sqrt{\lambda} h$, so that \eqref{result} becomes
\be\label{result2}
 Z_{\lambda}[f,J] = \int D h[x] \,  e^{i  \int \d^2 x   \left[ \frac{1}{2 } \epsilon^{\mu\nu} \epsilon_{ab} h_{\mu}^a h_{\nu}^b  \right]}  Z_{0}[f+\sqrt{\lambda} h,J] \, .
\ee
Hence provided the gaussian path integral is normalized as
\be
\int D h[x] \,  e^{i  \int \d^2 x   \left[ \frac{1}{2 } \epsilon^{\mu\nu} \epsilon_{ab} h_{\mu}^a h_{\nu}^b  \right]}  =1 \, ,
\ee
we satisfy the requirement that
\be
\lim_{\lambda \rightarrow 0}  Z_{\lambda}[f,J] = Z_{0}[f,J] \, .
\ee

The stochastic type path integral \eqref{stochasticpath} is closely similar to the holographic bulk cutoff picture \cite{McGough:2016lol} where $\lambda$ is playing the role of the holographic extra dimension. Indeed, the derivation of \eqref{result} is closely similar in spirit to the derivation of the solution of the Wheeler-de Witt equation given in \cite{Freidel:2008sh} as emphasized in \cite{McGough:2016lol}. It is worth noting that it would be straightforward to include curvature terms in \eqref{functionalequation}, in the path integral solution \eqref{stochasticpath}. The result of this would be that we no longer have $\frac{\partial P[\lambda]}{\partial \lambda} =0$, which will prevent us from giving a simple two dimensional path integral result.

\subsection{Conformal Field Theories}\label{CFT}

Up until now we have many no assumption about the matter Lagrangian $S_{0}[\varphi,e]$. We now assume that the action is classically conformally invariant $S_{0}[\varphi,e]=S_{CFT}[\varphi,e]$ for which $S_{CFT}[\{ \Omega^{-\Delta_I}\varphi_I \}, \Omega e]=S_{CFT}[\{\varphi_I \},e]$, with $\Delta_I$ the conformal weight of $\varphi_I$. The mass term breaks the conformal symmetry, but it can be reintroduced by means of a conformal \stu field $\Omega$\footnote{Often known as conformal compensator in the supergravity literature.} by performing the replacement
\be
e \rightarrow  \hat \Omega e \, ,  \quad g \rightarrow \hat \Omega^2 g \, , \quad \varphi_I \rightarrow  \hat \Omega^{-\Delta_I}\varphi_I  \, .
\ee
With the addition of the conformal compensator we have the manifestly conformally invariant action
\ba
&& S_{T\bar T}[\varphi,\gamma,\hat \Omega,e] =S_{CFT}[\varphi,e] +   \\
&& \int \d^2 x  \,  m^2  \( \hat \Omega^2  \sqrt{-\det g} +  \sqrt{-\det \gamma} - \hat \Omega\sqrt{ -\det g \tr[g^{-1} \gamma]  + 2 \sqrt{-\det g}\sqrt{-\det( \gamma)} } \) \, . \nn
\ea
The conformal \stu field arises as an auxiliary variable and so can be directly integrated out.  The equation for the conformal \stu field is
\be
\hat \Omega = \frac{1}{2} \sqrt{  \tr[g^{-1} \gamma]  + 2 \sqrt{\frac{-\det  \gamma}{-\det g}} }  \, ,
\ee
and so on substituting back in the action we have
\be
S_{T\bar T}[\varphi,\gamma,e] = \int \d^2 x  \,  \frac{m^2}{2}    \sqrt{-\det \gamma} - \frac{m^2}{4} \sqrt{-g} \(  \tr[g^{-1} \gamma]   \) + S_{CFT}[\varphi,e] \, .
\ee
or more explicitly in terms of the diffeomorphism \stu fields
\be\label{mainresult}
S_{T\bar T}[\varphi,\gamma,e] = \int \d^2 x  \,  \frac{m^2}{2}    \sqrt{-\det \partial_{\mu} \Phi^A \partial_{\nu} \Phi^B \hat \gamma_{AB}( \Phi)} - \frac{m^2}{4} \sqrt{-g}  g^{\mu\nu} \partial_{\mu} \Phi^A \partial_{\nu} \Phi^B \hat \gamma_{AB}( \Phi)+ S_{CFT}[\varphi,e] \, .
\ee
This is our central result. The complicated square root structure of the general massive gravity Lagrangian has disappeared, leaving behind standard sigma model Lagrangians. This result is remarkable in its simplicity and would have been hard to anticipate from the outset. \\

The first term in \eqref{mainresult} is effectively a total derivative and does not lead to any contribution to the equations of motion for $\Phi^A$. Indeed in the absence of boundary term considerations we have by means of a field dependent diffeomorphism
\be
\int \d^2 x  \,  \frac{m^2}{2}    \sqrt{-\det \partial_{\mu} \Phi^A \partial_{\nu} \Phi^B \hat \gamma_{AB}( \Phi)} =\int \d^2 x  \,  \frac{m^2}{2}    \sqrt{-\det  \gamma_{\mu\nu}( x)} \, , 
\ee
which is a field independent constant. We will see in section \ref{Bfield} that it is equivalent to a $B$-field interaction. Thus the relevant part of the action, at least for the local dynamics of the \stu and matter fields is
\be\label{Polyakov}
S_{P}[\varphi,\gamma,e] = \int \d^2 x  \, \[  - \frac{m^2}{4} \sqrt{-g}  g^{\mu\nu} \partial_{\mu} \Phi^A \partial_{\nu} \Phi^B \hat \gamma_{AB}( \Phi) \]+ S_{CFT}[\varphi,e] \, .
\ee
This is the classical action for a (in general) non-critical string model in Brink, Di Vecchia, Howe, Deser, Zumino form \cite{Brink:1976sc,Deser:1976rb}, usually referred to as Polyakov form.
We thus conclude that the \stu fields $\Phi^A$ act as two additional massless scalars with associated curved target space geometry $\gamma_{AB}( \Phi)$.  \\

For instance, if the original undeformed CFT had been a sum of $N$ scalar fields with moduli space metric $G_{IJ}(\varphi) $ on the curved spacetime $\gamma$
\be
S_{CFT}[\varphi,f]=\int \d^2 x \,  \sqrt{-\det \gamma} \(-\frac{1}{2} G_{IJ}(\varphi)\gamma^{\mu\nu}  \partial_{\mu}  \varphi^I \partial_{\nu} \varphi^I  \) \, ,
\ee
then the deformed theory is 
\be
S_{P}[\varphi,\gamma,e] = \int \d^2 x  \, \[  - \frac{m^2}{4} \sqrt{-g}  g^{\mu\nu} \partial_{\mu} \Phi^A \partial_{\nu} \Phi^B \hat \gamma_{AB}( \Phi)-\sqrt{-\det g} g^{\mu\nu} \frac{1}{2}  G_{IJ}(\varphi) \partial_{\mu} \varphi^I \partial_{\nu} \varphi^J  \] \, ,
\ee
which is a non-critical string theory with $N+2$ dimensional target space metric
\be
\d s^2_{\rm target} =  \frac{m^2}{2} \hat \gamma_{AB}( \Phi) \d \Phi^A \d \Phi^B+ G_{IJ}(\varphi) \d \varphi^I \d \varphi^J \, .
\ee
Equivalently the associated Nambu-Goto action is
\ba
S_{NG}&=&S_{T\bar T}=S_{P}+\int \d^2 x   \frac{m^2}{2}    \sqrt{-\det[ \partial_{\mu} \Phi^A \partial_{\nu} \Phi^B \hat \gamma_{AB}( \Phi)]}  \\
&=&\int \d^2 x   \frac{m^2}{2}    \sqrt{-\det[ \partial_{\mu} \Phi^A \partial_{\nu} \Phi^B \hat \gamma_{AB}( \Phi)]}- \sqrt{ - \det[ \frac{m^2}{2} \hat \gamma_{AB}( \Phi) \partial_{\mu} \Phi^A \partial_{\nu} \Phi^B+G_{IJ}(\varphi)  \partial_{\mu}  \varphi^I \partial_{\nu}  \varphi^J ]} \, , \nn
\ea
which more simply in unitary gauge $\Phi^a = x^a$ (often called static gauge) is
\ba
S_{NG}&=&\int \d^2 x \sqrt{-\det  \gamma}  \frac{m^2}{2}  \left[  1 - \sqrt{  \det[ \delta^{\mu}_{\nu}+\frac{2}{m^2} G_{IJ}(\varphi) \gamma^{\mu \alpha} \partial_{\alpha}  \varphi^I \partial_{\nu}  \varphi^J ]} \right]  \\
&=& \int \d^2 x \sqrt{-\det  \gamma}  \frac{m^2}{2}  \left[  1 - \sqrt{ 1+\frac{2}{m^2} G_{IJ}(\varphi) \gamma^{\mu \nu} \partial_{\mu}  \varphi^I \partial_{\nu}  \varphi^J +\frac{2}{m^4} D} \]  \\
&=& \int \d^2 x \sqrt{-\det  \gamma}  \frac{1}{2 \lambda}  \left[  1 - \sqrt{ 1+2 \lambda G_{IJ}(\varphi) \gamma^{\mu \nu} \partial_{\mu}  \varphi^I \partial_{\nu}  \varphi^J +2 \lambda^2 D} \]  \, ,
\ea
with
\be
D= \(G_{IJ}(\varphi) \gamma^{\mu \nu} \partial_{\mu}  \varphi^I \partial_{\nu}  \varphi^J  \)^2-  G_{IJ}(\varphi) \gamma^{\mu \alpha} \partial_{\alpha}  \varphi^I \partial_{\nu}  \varphi^J   G_{KL}(\varphi) \gamma^{\nu \beta} \partial_{\beta}  \varphi^K \partial_{\mu}  \varphi^L  \, .
\ee
We could of course have arrived at \eqref{mainresult} by first starting with the zweibein form of the action \eqref{actionTT}, introducing the conformal compensator $e \rightarrow \hat \Omega e$ and then integrating it out. This would result in the classically conformally invariant action
\be\label{alternativeconformal}
S_{T\bar T}[\varphi,\gamma,e]  = \int \d^2 x \, \[ m^2 \det f - \frac{1}{4} m^2 \frac{ \( \epsilon^{\mu \nu} \epsilon_{ab} e_{\mu}^a f_{\nu}^b\)^2}{\det e} \]+ S_{CFT}[\varphi,e]  \, .
\ee
Then integrating out the Lorentz \stu fields we would return to the metric formulation \eqref{mainresult}. There appears to be no particular advantage in working with \eqref{alternativeconformal}.

\subsection{Equivalence to Polyakov with a $B$-field}\label{Bfield}

An alternative representation of the $T\bar T$ action \eqref{mainresult} is obtained by recognizing that the first term in \eqref{mainresult}  is equivalent to turning on a $B$-field. The need to include a $B$-field coupling in flat spacetime had previously been noted in \cite{Callebaut:2019omt} and \cite{Hashimoto:2019wct}.
In order to see the equivalence, it is useful to use target spacetime diffeomorphisms to put the target spacetime metric in a conformally flat lightcone form
\be \label{conformallyflat}
\hat \gamma_{AB}(\Phi) \d \Phi^A \d \Phi^B  = 2 \hat \gamma_{+-}(\Phi) \d \Phi^+ \d \Phi^- \, .
\ee
Once this is done we recognize that (assuming $\hat \gamma_{+-}(\Phi)>0$)
\be
 \sqrt{-\det \( \partial_{\mu} \Phi^A \partial_{\nu} \Phi^B \hat \gamma_{AB}( \Phi)\)} = \hat \gamma_{+-}(\Phi)  \sqrt{-\det \( \partial_{\mu} \Phi^+ \partial_{\nu} \Phi^-+\partial_{\nu} \Phi^+ \partial_{\mu} \Phi^- \)} \, .
\ee
In two dimensions we have the identity (where $\epsilon^{\mu\nu}$ is the Levi-Civita symbol not tensor)
\be
\det \( \partial_{\mu} \Phi^+ \partial_{\nu} \Phi^-+\partial_{\nu} \Phi^+ \partial_{\mu} \Phi^- \)  = - \(\epsilon^{\mu\nu} \partial_{\mu} \Phi^+ \partial_{\nu} \Phi^- \)^2 \, ,
\ee
and so further assuming $\epsilon^{\mu\nu} \partial_{\mu} \Phi^+ \partial_{\nu} \Phi^- \le 0$ we have
\be
\sqrt{-\det \( \partial_{\mu} \Phi^A \partial_{\nu} \Phi^B \hat \gamma_{AB}( \Phi)\)} =-\hat \gamma_{+-}(\Phi) \epsilon^{\mu\nu} \partial_{\mu} \Phi^+ \partial_{\nu} \Phi^- = -\frac{1}{2 }B_{AB}(\Phi) \epsilon^{\mu\nu} \partial_{\mu} \Phi^A \partial_{\nu} \Phi^B \, .
\ee
Here we have defined an antisymmetric $B$-field with components in conformally flat gauge $B_{+-}(\Phi)=\hat\gamma_{+-}(\Phi)$.
Hence the $T\bar T$ action \eqref{mainresult} is simply that of a Polyakov string in a non-zero $B$-field
\be\label{mainresult2}
S_{T\bar T}[\varphi,\gamma,e] = \int \d^2 x  \,  - \frac{m^2}{4} \left[ \sqrt{-g}  g^{\mu\nu} \partial_{\mu} \Phi^A \partial_{\nu} \Phi^B \hat \gamma_{AB}( \Phi) + B_{AB}(\Phi) \epsilon^{\mu\nu} \partial_{\mu} \Phi^A \partial_{\nu} \Phi^B \right] + S_{CFT}[\varphi,e] \, .
\ee
In the flat space limit this reproduces the observation of \cite{Callebaut:2019omt} that a constant $B$ field of magnitude $1$, i.e. $B_{AB}=\epsilon_{AB}$ is needed.

\subsection{Special Non-Conformal Field Theories}\label{specialconformal} 

We now consider a special subclass of non-conformal field theories for which the action takes the form
\be
S_0[\varphi,e] =S_{CFT}[\varphi,e] -\int \d^2 x \det e \, V[\varphi] \, ,
\ee
where $ V[\varphi] $ is conformally invariant, so that on introducing the conformal \stu field  $e \rightarrow \hat \Omega e$, $\varphi_I \rightarrow  \hat \Omega^{-\Delta_I}\varphi_I  $
\be
S_0[ \{\hat \Omega^{-\Delta_I}\varphi_I \}, \hat \Omega e] =S_{CFT}[\varphi,e] -\int \d^2 x \det e \, \hat \Omega^2 \, V[\varphi] \, .
\ee
These describe for example theories of minimally coupled scalars with a potential $V[\varphi] $ such as the sine-Gordon model. The key virtue of this special class is that the equation for the conformal compensator remains quadratic, with solution now
\be
\hat \Omega = \frac{1}{(1-V[\varphi]/m^2)}\frac{1}{2} \sqrt{  \tr[g^{-1} \gamma]  + 2 \sqrt{\frac{-\det \gamma}{-\det g}} }  \, .
\ee
Substituting back in we find
\ba\label{otherform}
S_{T\bar T}[\varphi,\gamma,e] &=& \int \d^2 x  \,  \frac{m^2}{2}   \frac{(1-2 V[\varphi]/m^2)}{(1-V[\varphi]/m^2)} \sqrt{-\det \partial_{\mu} \Phi^A \partial_{\nu} \Phi^B \hat \gamma_{AB}( \Phi)} \nn \\
&& - \frac{m^2}{4(1-V[\varphi]/m^2)} \sqrt{-g}  g^{\mu\nu} \partial_{\mu} \Phi^A \partial_{\nu} \Phi^B \hat \gamma_{AB}( \Phi)+ S_{CFT}[\varphi,e] \, .
\ea
This takes a Polyakov form with the addition of a non-standard potential for the matter fields. 
Following section \ref{Bfield} this is equivalent to a Polyakov action with a slightly modified target spacetime metric and with a modified non-zero $B$-field
\be\label{otherform2}
S_{T\bar T}[\varphi,\gamma,e] =\int \d^2 x  \,  - \frac{m^2}{4} \left[ \sqrt{-g}  g^{\mu\nu} \partial_{\mu} \Phi^A \partial_{\nu} \Phi^B \hat \gamma'_{AB}( \varphi,\Phi) + B'_{AB}(\varphi,\Phi) \epsilon^{\mu\nu} \partial_{\mu} \Phi^A \partial_{\nu} \Phi^B \right] + S_{CFT}[\varphi,e] \, .
\ee
with 
\be
\hat \gamma'_{AB}( \varphi,\Phi) =\frac{1}{(1-V[\varphi]/m^2)}\hat \gamma_{AB}( \Phi) \, , \quad \text{and  }  B'_{AB}(\varphi,\Phi) = \frac{(1-2 V[\varphi]/m^2)}{(1-V[\varphi]/m^2)}B_{AB}(\Phi)  \, .
\ee
where $B_{AB}(\Phi) $ is the same $B$-field as needed for $V[\varphi]=0$, namely $B_{+-}(\Phi)=\hat\gamma_{+-}(\Phi)$ in conformally flat gauge \eqref{conformallyflat}.

\subsection{Field Dependent Diffeomorphisms}\label{Fielddiffs}

Since any metric in two dimensions is locally conformally flat, we may always locally write  $F_{A}^a(\Phi) = e^{2 \beta(\Phi)} \delta_{A}^a $, 
so that the reference zweibein takes the form
\be
f_{\mu}^a= e^{2 \beta(\Phi)}  \Lambda^a{}_A \partial_{\mu} \Phi^A \, .
\ee
This representation is only globally well-defined for two dimensional spacetimes which are topolgically $R^{1,1}$. Working with the vielbein form of the massive gravity action \eqref{actionTT} it is straightforward to see that the equation of motion for the $\Phi^A$ is entirely determined by the term linear in $f$, namely
\be
 \int \d^2 x  \,  \(  - m^2  \epsilon^{\mu\nu} \epsilon_{ab} e_{\mu}^a f_{\nu}^b \) \, ,
\ee
since in the term quadratic in $f$ we may always perform a diffeomorphism that removes the $\Phi^a$ dependence. Working in the local Lorentz gauge where $\Lambda^a{}_A =1$ then the equation of motion is
\be\label{condition1}
- \partial_{\nu} \( e^{2 \beta(\Phi)}  \epsilon^{\mu\nu} \epsilon_{ab} e_{\mu}^a  \)+ 2 e^{2 \beta(\Phi)} \frac{\partial \beta}{\partial \Phi^b} \epsilon^{\mu\nu} \epsilon_{ac}  e^{a}_{\mu} \partial_{\nu} \Phi^c =0 \, ,
\ee
In general this is complicated to solve, but if the reference metric is Minkowski, i.e. if $\beta=0$ it reduces to
\be \label{condition2}
\partial_{\nu} \(  \epsilon^{\mu\nu} \epsilon_{ab} e_{\mu}^a  \)=0
\ee
which is easily solved by $e^a_{\mu}(x) = \partial_{\mu} Z^a(x)$, hence the metric $g$ is itself also Minkowski. Even more remarkably we find that the local Lorentz transformations that describe the zweibein for each Minkowski metric are tied to each other, so that in the gauge in which the Lorentz \stu fields for $f$ are zero, then those for $e$ are zero. The equations of motion for matter fields $\varphi$ are the standard ones in the unitary gauge for which $Z^a = x^a$. Hence the solutions in the unitary gauge $\Phi^a=x^a$ can be inferred by means of a field dependent diffeomorphism. The $Z^a(x)$ are the dual \stu fields, and arise in the same manner as they do in bigravity theories \cite{Fasiello:2013woa}. The two sets of \stu fields allows us to give two distinct descriptions of the same physics, which are related to each other by a field dependent diffeomorphism. In higher dimensions, the Galileon duality transformations  \cite{Fasiello:2013woa,Curtright:2012gx,deRham:2013hsa,deRham:2014lqa} are special case of the equivalence of the two descriptions. 
For instance as a special case of the duality transformations, we make work in more general unitary gauge defined by
$\Phi^a = x^a+ s \Pi^a$ and $Z^a=x^a+(1-s) \Pi^a$. This will give rise to a one-parameter family of equivalent descriptions of the same physics, fields at different values of $s$ related to each other by non-local transformations. \\

It terms of solving the system, it is most useful to work in the unitary gauge with $Z^a = x^a$, i.e. for which $e^a_{\mu} = \delta^a_{\mu}$ then \eqref{eom1} turns into an equation for $\Phi^a$ which is simply
\be
\partial_{\mu} \Phi^a = \delta_{\mu}^a + \lambda \epsilon_{\mu\nu} \epsilon^{ab} T_0 ^{\nu}{}_b \, ,
\ee
where $T_0 ^{\nu}{}_n $ is the Minkowski space stress energy of the undeformed theory. That this equation is integrable follows from the fact that
\be
\epsilon^{\omega \mu} \partial_{\omega}\partial_{\mu} \Phi^a=0 =  \lambda  \epsilon^{\omega \mu} \epsilon_{\mu\nu} \epsilon^{ab} \partial_{\omega}T_0 ^{\nu}{}_b  = - \lambda  \partial_{\nu}T_0 ^{\nu}{}_b=0\, ,
\ee
by virtue of conservation of stress energy on a Minkowski background. These relations are extremely useful, and have for example been used in \cite{Dubovsky:2017cnj} to provide a derivation of the CDD factors for the S-matrix from Jackiw-Teitelboim gravity and in for example \cite{Conti:2018tca,Coleman:2019dvf} to provide a means of constructing solutions in $T \bar T$ deformed theories.\\

In the more general curved case, we can make the ansatz
\be
e_{\mu}^a= e^{2 \tilde \beta(\Phi)}  \tilde \Lambda^a{}_B \partial_{\mu} Z^B \, .
\ee
The Lorentz \stu fields should be determined by symmetric vielbein condition \eqref{symmetriczweibein} which in the Lorentz gauge for which $ \Lambda^a{}_B=\delta_a^B$ gives
\be
\epsilon^{\mu\nu} e_{\mu}^a f_{\nu}^b {\eta_{ab}}=0 \rightarrow \epsilon^{\mu\nu}  \tilde \Lambda^a{}_B \partial_{\mu} Z^B \partial_{\nu} \Phi^b  {\eta_{ab}}=0 \, .
\ee
For a curved reference metric, $\tilde  \Lambda^a{}_B= \delta_a^B$ is no longer the solution since in general $\epsilon^{\mu\nu}  \partial_{\mu} Z^a \partial_{\nu} \Phi^b  {\eta_{ab}} \neq 0$. The same is true in Minkowski, but there the additional requirement \eqref{condition2} enforces $\tilde  \Lambda^a{}_B= \delta_a^B$. By contrast \eqref{condition1} becomes 
\be\label{condition3}
- \partial_{\nu} \( e^{2 \beta(\Phi)+2 \tilde \beta(\Phi)}  \epsilon^{\mu\nu} \epsilon_{ab}  \tilde \Lambda^a{}_B \partial_{\mu} Z^B  \)+ 2 e^{2 \beta(\Phi)+2 \tilde \beta(\Phi)} \frac{\partial \beta}{\partial \Phi^b} \epsilon^{\mu\nu} \epsilon_{ac}  \tilde \Lambda^a{}_B \partial_{\mu} Z^B \partial_{\nu} \Phi^c =0 \, .
\ee
which even in the unitary gauge $\Phi^a=x^a$ does not appear to simplify in any straightforward way, even if we focus on its antisymmetric part. Hence in a curved spacetime we expect a non-trivial local Lorentz transformation and conformal transformation between the two zweibeins. The solution for the Lorentz transformations can be obtained analogous to \eqref{squareroot} in terms of the characteristic square root structure. However it does mean that the special features that make the $T \bar T$ deformation trivially solvable on Minkowski spacetime do not automatically extend to curved spacetime.

\section{From Polyakov to $T \bar T$}\label{PolyakovTTbar}

The simplicity of the Polyakov form of the action \eqref{Polyakov} suggests that we should have been able to derive it from the outset, without needing to pass through the massive gravity construction, at least in the case where the seed theory is a CFT. Indeed starting with a non-critical string theory, we may infer the correspondence to a $T\bar T$ deformed theory as we outline below. \\

Consider a classical non-critical string action in Polyakov form, and separate out two of the target space dimensions. We define by $(2 \lambda)^{-1}\hat \gamma_{AB}$ the two dimensional metric associated with these two target space dimensions assuming that an overall scale $\lambda$ can be factored out. In the case of a curved metric $1/\sqrt{\lambda}$ is essentially the overall curvature scale. In the case of a flat toroidal metric it is essentially the size of the torus. 

Classically the non-critical string action can then be split up as
\be\label{Poly}
S_{P}[\varphi,\gamma,e] = \int \d^2 x  \, \[  -\frac{1}{4 \lambda} \sqrt{-\det g}  g^{\mu\nu} \hat \gamma_{AB}(\Phi) \partial_{\mu} \Phi^A \partial_{\nu} \Phi^B \]+ S_{CFT}[\varphi,e] \, .
\ee
The string tension has been absorbed in a canonical normalization of the fields $\Phi^a$. Here $g$($e$) is implicitly a function of $\lambda$ by virtue of its on-shell equation of motion. We then have up to terms which vanish on-shell
\be
\frac{\d}{\d \lambda} S_{P}[\varphi,\gamma,e]  = \int \d^2 x  \,  \frac{1}{4 \lambda^2} \sqrt{-\det g}  g^{\mu\nu} \gamma_{\mu\nu} \, ,
\ee
where we use the shorthand $\gamma_{\mu\nu} = \hat \gamma_{AB}(\Phi) \partial_{\mu} \Phi^A \partial_{\nu} \Phi^B$.
Associated with the two dimensional sub-manifold of the target space with metric $\hat \gamma_{AB}(\Phi)$, we may define the stress energy tensor
\be
\sqrt{-\det{\hat \gamma}} {\hat T}^{AB}(\Phi) = 2 \frac{\delta }{\delta \hat \gamma_{AB}(\Phi)} S_{P}[\varphi,\gamma,e] +\frac{1}{2\lambda} \sqrt{-\det{\hat \gamma}}  \hat \gamma^{AB}(\Phi) \, ,
\ee
where the functional variation is defined in the sense
\be
\delta S_P = \int \d^2 \Phi \, \frac{\delta }{\delta \hat \gamma_{AB}(\Phi)} S_{P}[\varphi,\gamma,e]  \delta \hat \gamma_{AB}(\Phi) \, .
\ee
This is the naive stress energy tensor of the non-critical string action defined relative to the target space metric, shifted by a cosmological constant term whose relevance is clear with hindsight. It proves slightly more convenient to rewrite this in terms of a stress energy associated with the induced metric $\gamma_{\mu\nu}$ via
\be
{\hat T}^{AB} = \partial_{\mu} \Phi^A \partial_{\nu} \Phi^B T^{\mu\nu} \, ,
\ee
where equivalently
\ba
\sqrt{-\det{\gamma}} T^{\mu\nu}(x) &=& \frac{1}{2\lambda}\sqrt{-\det{\gamma}}  \gamma^{\mu\nu}(x)+ 2 \frac{\delta }{\delta \gamma_{\mu \nu}(x)} S_{P}[\varphi,\gamma,e] \label{Tdef1} \\
&=&  \frac{1}{2\lambda}\sqrt{-\det{\gamma}}  \gamma^{\mu\nu}(x)-\frac{1}{2\lambda}\sqrt{-\det{g}}  g^{\mu\nu}(x) \, .
\ea
That is
\be\label{Tdef2}
T^{\mu\nu}(x) =\frac{1}{2\lambda} \gamma^{\mu\nu}(x)-\frac{1}{2\lambda} \frac{\sqrt{-\det{g}}}{\sqrt{-\det{\gamma}} }  g^{\mu\nu}(x)  \, .
\ee
This equation is the analogue of \eqref{efrelation}.  We note that $\gamma^{\mu\nu}$ are the components of the inverse of $\gamma_{\mu\nu}$, i.e. $\gamma^{\mu\nu}\gamma_{\nu\rho} =\delta^{\mu}_{\rho}$, so that the indices on it are not being raised with $g^{\mu\nu}$.  In particular if we lower one index we have
\be\label{Tdef3}
T^{\mu}{}_{\nu}(x) =\frac{1}{2\lambda} \delta^{\mu}_{\nu}-\frac{1}{2\lambda} \frac{\sqrt{-\det{g}}}{\sqrt{-\det{\gamma}} }  g^{\mu\alpha}(x) \gamma_{\alpha \nu}  \, .
\ee
From \eqref{Tdef2} with a little rearrangement we find
\be
\det \[\sqrt{-\det g} g^{\mu\nu}\]=-1 = \det\[  \sqrt{-\det \gamma}( \gamma^{\mu\nu} - 2\lambda T^{\mu\nu} ) \right] = - \det\[ \gamma_{\mu}{}^{\nu} -2 \lambda T_{\mu}{}^{\nu} \] \, ,
\ee
which implies the classical trace flow equation
\be\label{traceflow}
\Theta = 2 \lambda \det[T_{\mu}^{\nu}] = -\lambda ( T_{\mu\nu} T^{\mu\nu} - \Theta^2 ) \, ,
\ee
where we have defined the trace of the stress energy tensor again as $T^{\mu\nu}(x) \gamma_{\mu\nu} = \Theta$. 

From the perspective of the non-critical string, \eqref{Tdef2} is an unusual (composite) operator to define since it is not immediately clear it is associated with the covariant stress energy of the full CFT, neither that of $S_{CFT}[\varphi,e] $.
The covariant stress energy tensor is
\ba
&& \sqrt{-\det g} T^{\mu\nu}_{\rm cov}  = 2 \frac{\delta }{\delta g_{\mu \nu}(x)} S_{P}[\varphi,\gamma,e] \\
&=&\frac{1}{2 \lambda} \sqrt{-\det g} \(g^{\mu \alpha} g^{\nu\beta} \gamma_{\alpha \beta} - \frac{1}{2} g^{\mu\nu} g^{\alpha \beta} \gamma_{\alpha \beta} \)  +\sqrt{-\det g} T^{\mu\nu}_{CFT}  \, ,
\ea
where $\sqrt{-g}  T^{\mu\nu}_{CFT} (x)= 2 \frac{\delta }{\delta g_{\mu \nu}(x)} S_{CFT}[\varphi,e]$. The vanishing of the covariant stress energy tensor $T^{\mu\nu}_{\rm cov} =0$ implies
\be\label{gammagT}
\gamma_{\mu\nu} =\frac{1}{2} \gamma_{\alpha \beta} g^{\alpha \beta}   g_{\mu\nu} -  2\lambda T^{CFT}_{\mu\nu}  \, .
\ee
where $T^{CFT}_{\mu\nu} = g_{\mu \alpha} g_{\nu \alpha} T_{CFT}^{\alpha \beta}$. This is the analogue of \eqref{efrelation2}. The combination $\gamma_{\alpha \beta} g^{\alpha \beta} $ cannot be determined from this equation by virtue of the underlying conformal symmetry, i.e. that $g_{\alpha \beta}T_{CFT}^{\alpha \beta}=0$.
Rewriting \eqref{gammagT} we have
\be
g_{\mu\nu} = \alpha ( \gamma_{\mu\nu} + 2\lambda T^{CFT}_{\mu\nu} ) \, ,
\ee
with $\alpha$ an undetermined conformal factor. Substituting into \eqref{Tdef2} gives the relation between the two stress energies
\ba
T^{\mu \nu} &=&\frac{1}{2\lambda} \gamma^{\mu \nu}-\frac{1}{2\lambda} \sqrt{\det(1+2\lambda \gamma^{-1} T^{CFT}    )}  [( \gamma + 2\lambda T^{CFT} )^{-1}]^{\mu\nu} \\
&=& T_{CFT}^{\mu\nu} + {\cal O}(\lambda) \, .
\ea
This equation is the direct analogue of \eqref{TTrelation} and its form explains why we included a cosmological constant term in the definition of the stress energy tensor. We see that $T^{\mu \nu}$ is a $\lambda$ deformed version of the stress tensor of the original CFT $S_{CFT}[\varphi,e]$.

Making use of the trace of \eqref{Tdef2} we have
\be
\Theta = T^{\mu\nu} \gamma_{\mu\nu} =\frac{1}{\lambda}- \frac{1}{2 \lambda} \frac{\sqrt{-\det g}}{\sqrt{-\det \gamma}} g^{\mu\nu} \gamma_{\mu\nu} \, .
\ee
so that the flow equation is
\be
\frac{\d}{\d \lambda} S_{P}[\varphi,\gamma,e] =-\int \d^2 x  \[  \,  \frac{1}{2 \lambda}\sqrt{-\det{\gamma}} \Theta- \frac{1}{2 \lambda^2} \sqrt{-\det \gamma} \] \, .
\ee
Equivalently defining the cosmological constant shifted action
\be\label{actionshift}
S_{\lambda}[\varphi,\gamma]  =  S_{P}[\varphi,\gamma,e_*] + \int \d^2 x \frac{1}{2 \lambda} \sqrt{-\det \gamma} \, ,
\ee
then
\be
\frac{\d}{\d \lambda} S_{\lambda}[\varphi,\gamma] =-\int \d^2 x  \[  \,  \frac{1}{2 \lambda}\sqrt{-\det{\gamma}} \Theta\] \, .
\ee
This relation appears to be very different that \eqref{TTdef}. Indeed, it cannot be used to define a perturbative expansion in $\lambda$ since the right hand side scales like $1/\lambda$. The naive divergence is of course cancelled by the fact that classically $\Theta_{\lambda=0}=0$, but this implies we already need to know the first order perturbation of the stress energy on the RHS in order to determine the first order perturbation of the action! 

Fortunately this is easily dealt with. Making use of the trace flow equation \eqref{traceflow} we have
\be\label{TTbar3}
\frac{\d}{\d \lambda} S_{\lambda}[\varphi,\gamma] =\frac{1}{2} \int \d^2 x  \sqrt{-\det{\gamma}}  ( T_{\mu\nu} T^{\mu\nu} - \Theta^2 ) \, .
\ee
which can now be used to define the deformation perturbatively. This is exactly the metric version of the $T\bar T$ deformation \eqref{TTdef} given that \eqref{Tdef1} implies that $T^{\mu\nu}$ is indeed the appropriately defined stress energy
\be
\sqrt{-\det{\gamma}} T^{\mu\nu}(x) =  2 \frac{\delta }{\delta \gamma_{\mu \nu}(x)} S_{\lambda}[\varphi,\gamma] \, .
\ee
Thus there is a one to one correspondence between the $T\bar T$ deformed action $S_{\lambda}[\varphi,\gamma]$ and the Polyakov form non-critical string action $S_{P}[\varphi,\gamma,e]$, for a seed CFT, the two being related by the target space cosmological constant shift \eqref{actionshift}. Since for a fixed reference metric the addition of a cosmological constant to the action is irrelevant, we conclude that classically the non-critical string theory described by the Polyakov action \eqref{Poly} is identical to the $T\bar T$ deformed theory defined by \eqref{TTbar3}. However we may do better than that. Following the observations of section \ref{Bfield}, the two dimensional cosmological constant term may be equivalently rewritten as a $B$-field contribution and so we recognize that the $T \bar T$ deformed action $S_{\lambda}[\varphi,\gamma] $ is simply the Polyakov action with the addition of a specific non-zero $B$-field \eqref{mainresult2}.

\subsection{$T \bar T$ Hamiltonian for a CFT}\label{CFTHamiltonian}

Remarkably we will now see that it is trivial to construct the Hamiltonian of the $T \bar T$ deformation for any CFT without needing to solve any differential equation or even algebraic relations. As in section \ref{HamiltonianQFT} we may always choose to parameterize the induced target space metric in ADM form as 
\be
\gamma_{\mu\nu} \d x^{\mu} \d x^{\nu} = - M^2(x,t)  \d t^2 + B^2(x,t)  ( \d x+ M^x(x,t)  \d t)^2 \, .
\ee
where in this subsection we work in unitary gauge $\Phi^a = x^a$. Since the Polyakov action is conformally invariant, we can without any loss of generality fix the dynamical metric in the ADM form
\be
g_{\mu\nu} \d x^{\mu} \d x^{\nu}  = - N^2(x,t) \d t^2 +  B^2(x,t) ( \d x+ N^x(x,t) \d t)^2 \, .
\ee
In doing so we note that only $N$ and $N^x$ are dynamical, and the spatial scale factor is fixed using conformal symmetry to be the same as the target space metric. This means that $g_{\mu\nu} = \gamma_{\mu\nu}$ when $N=M$ and $N^x=M^x$.  We write the action for the undeformed theory in the same canonical form as considered in section \ref{HamiltonianQFT}
\be\label{canonicalaction}
 S_{CFT}[\varphi,e] = \int \d^2 x \left[  \sum_I \Pi_I \dot \varphi_I - N {\cal H} - N^x {\cal H}_x- \sum_{a} \lambda_a C_a\right] \, .
\ee
The stress energy defined relative to the $\gamma$ metric is from \eqref{Tdef2}
\be
T^{\mu}{}_{\nu} = \frac{1}{2\lambda} \delta^{\mu}_{\nu} - \frac{1}{2\lambda} \frac{\sqrt{-\det g}}{\sqrt{-\det \gamma}} g^{\mu \alpha} \gamma_{\alpha \nu} \, ,
\ee
which can be conveniently denoted as
\be
T^{\mu}{}_{\nu} = \frac{1}{2\lambda}\begin{bmatrix} 
1-\( \frac{1}{N M}(M^2-B^2 {M^x}^2)+ \frac{B^2M^x N^x}{N M} \)&- \frac{B^2 (N^x-M^x)}{N M} \\
-\( \frac{M^x N}{M}- \frac{N^x M}{N} - \frac{B^2 N^x M^x(N^x-M^x)}{N M}\)  &1- \( \frac{B^2 M^x N^x}{N M}+\frac{N}{M}- \frac{B^2 {N^x}^2}{N M} \)
\end{bmatrix}\, .
\ee
Furthermore
\be
Tr[g^{-1} \gamma] = 1+ \frac{M^2}{N^2}- \frac{B^2}{N^2} (M^x-N^x)^2 \, .
\ee
Then the deformed action using the Polyakov action is then
\be
S_{\lambda} =  \int \d^2 x \left[ \frac{1}{2\lambda}M B - \frac{N B }{4 \lambda} \(1+ \frac{M^2}{N^2}- \frac{B^2}{N^2} (N^x-M^x)^2 \)+ \sum_I \Pi_I \dot \varphi_I - N {\cal H} - N^x {\cal H}_x - \sum_{a} \lambda_a C_a\right] \, .
\ee
Varying with respect to $N^x$ gives $N^x = M^x +2 \lambda N {\cal H}_x/B^3 $ and substituting back in 
\be
S_{\lambda} =  \int \d^2 x \left[  \frac{1}{2\lambda}M B - \frac{NB}{4 \lambda} \(1+ \frac{M^2}{N^2}\ \)- \frac{N \lambda {\cal H}_x^2}{B^3} +  \sum_I \Pi_I \dot \varphi_I - N {\cal H} - M^x {\cal H}_x- \sum_{a} \lambda_a C_a \right] \, .
\ee
In turn, varying with respect to $N$ then gives
\be
N = \frac{M}{\sqrt{1+ \frac{4 \lambda}{B} \({\cal H} + \frac{\lambda {\cal H}_x^2}{ B^3} \) }} \, ,
\ee
and so substituting back in we have the final $T \bar T$ deformed action defined on an arbitrary curved spacetime
\be\label{finalaction}
S_{\lambda} =  \int \d^2 x \left[  \sum_I \Pi_I \dot \varphi_I - \frac{M B}{2\lambda}\( -1 + \sqrt{1+  \frac{4 \lambda}{B} \({\cal H} + \frac{ \lambda {\cal H}_x^2}{ B^3} \)} \)- M^x {\cal H}_x- \sum_{a} \lambda_a C_a \right] \, .
\ee
Once again this result is remarkable in its simplicity. Equivalently the Hamiltonian defined on the constraint surface $C_a=0$ is
\be
H= \int \d x \frac{M B}{2\lambda}\( -1 + \sqrt{1+  \frac{4 \lambda}{B} \({\cal H} + \frac{ \lambda {\cal H}_x^2}{ B^3} \)} \)+ M^x {\cal H}_x \,.
\ee
In particular working in the coordinate system where $M=1$ and $M^x=0$ we find
\be\label{HamiltonianCFT}
H= \int \d x \frac{B}{2\lambda}\( -1 + \sqrt{1+  \frac{4 \lambda}{B} \({\cal H} + \frac{ \lambda {\cal H}_x^2}{ B^3} \)} \) \, .
\ee

\paragraph{Burgers Equation:} This square root structure is exactly the form we obtain from solving the inviscid Burgers equation
\be
 \partial_{\lambda} E_n(R,\lambda) = E_n(R,\lambda) \partial_R E_n(R,\lambda)  + \frac{1}{R} P_n(R)^2 \, ,
\ee
with initial conditions appropriate for a CFT on a cylinder of radius $R$
\be
E_n(R,0) = (n+\bar n- c/12)/R \, , \quad P_N(R) =(n-\bar n)/R \, ,
\ee
which gives
\be\label{Burgerssoln}
E_n(R,\lambda) = \frac{R}{2 \lambda} \(-1 + \sqrt{1+ \frac{4 \lambda E_n(R,0)}{R} + \frac{4 \lambda^2 P_n^2}{R^2}} \) \, .
\ee
Indeed if we take the reference metric to be flat and compactify $x \in [0,1]$ and take $B=R$ then \eqref{HamiltonianCFT} is precisely \eqref{Burgerssoln} given the identification
\be
E_n(R,\lambda) = \langle n | H | n \rangle ,  \, \quad E_n(R,0)= \langle n | {\cal H} | n \rangle \, \quad P_n(R) =\frac{1}{R} \langle n | {\cal H}_x | n \rangle \, ,
\ee
for an energy eigenstate labelled by $(n,\bar n)$. Equation \eqref{HamiltonianCFT} is the appropriate classical generalization to a curved spacetime. The universality of this form is remarkable, although as we have seen is directly tied to the assumed conformal invariance of the seed theory. \\

To confirm that \eqref{finalaction} is the correct $T \bar T$ deformation for a general CFT, we differentiate
\be
\frac{\d S_{\lambda} }{\d \lambda} = \int \d^2 x \, MB \frac{\(1+ 2\lambda \frac{{\cal H}}{B} - \sqrt{1+  \frac{4 \lambda}{B} \({\cal H} + \frac{ \lambda {\cal H}_x^2}{ B^3} \)}  \)}{2\lambda^2 \sqrt{1+  \frac{4 \lambda}{B} \({\cal H} + \frac{ \lambda {\cal H}_x^2}{ B^3} \)} } \, .
\ee
The trace of the stress energy tensor is explicitly
\ba
\Theta &=& \frac{1}{\lambda} - \frac{N}{2M \lambda} Tr[g^{-1} \gamma] = \frac{1}{2\lambda} \( 2 - \frac{N}{M}-\frac{M}{N}+\frac{B^2}{N M} (M^x-N^x)^2 \)  \, ,\nn \\
&=& -  \frac{\(1+ 2\lambda \frac{{\cal H}}{B} - \sqrt{1+  \frac{4 \lambda}{B} \({\cal H} + \frac{ \lambda {\cal H}_x^2}{ B^3} \)}  \)}{\lambda \sqrt{1+  \frac{4 \lambda}{B} \({\cal H} + \frac{ \lambda {\cal H}_x^2}{ B^3} \)} } \, .
\ea
a result which may be derived directly by varying \eqref{finalaction} with respect to $\gamma$. Hence we conclude that
\be
\frac{\d S_{\lambda} }{\d \lambda} = -\int \d^2 x  \[  \,  \frac{1}{2 \lambda}\sqrt{-\det{\gamma}} \Theta\] =\int \d^2 x  \sqrt{-\det{\gamma}}  \frac{1}{2} ( T_{\mu\nu} T^{\mu\nu} - \Theta^2 )\, ,
\ee
as required.

\subsection{From Polyakov to Special Non-Conformal Field Theories}\label{specialclass}

Armed with the knowledge that the $T\bar T$ deformation is a direct consequence of the Polyakov action on a curved target space, we can easily imagine generalizations. For instance, from the perspective of the non-critical string, there is no reason why the target space metric should not include explicit dependence on $\varphi$ in front of $\gamma$, e.g. in the form
\be
S_{P}[\varphi,\gamma,e] = \int \d^2 x  \, \[  - \frac{1}{4} \sqrt{-\det g}  g^{\mu\nu}  e^{2 \chi(\varphi,\lambda)}\hat \gamma_{AB}(\Phi) \partial_{\mu} \Phi^A \partial_{\nu} \Phi^B \]+ S_{CFT}[\varphi,e] \, .
\ee
Differentiating with respect to $\lambda$ gives up to terms which vanish on-shell
\be
\frac{\d}{\d \lambda} S_{P}[\varphi,\gamma,e]  =- \int \d^2 x  \, \frac{1}{2} \frac{\partial \chi(\varphi,\lambda)}{\partial \lambda} \sqrt{-\det g}  e^{2 \chi(\Phi,\varphi,\lambda)} g^{\mu\nu} \gamma_{\mu\nu} \, ,
\ee
We define a closely related stress energy to earlier which comes from varying
\be\label{TTPrelationspecial}
S_{\bar T \bar T}=S_{P}[\varphi,\gamma,e] + \int \d^2 x \frac{1}{2} \alpha(\varphi,\lambda) \sqrt{-\det \gamma} \, ,
\ee
with respect to $\gamma$, namely (remember that indices of $T$ are lowered with $\gamma$)
\be
T^{\mu}{}_{\nu} =\frac{1}{2} \alpha(\varphi,\lambda)  \delta^{\mu}_{\nu}-\frac{1}{2} e^{2 \chi(\varphi,\lambda)}\frac{\sqrt{-\det{g}}}{\sqrt{-\det{\gamma}} }  g^{\mu\alpha}(x) \gamma_{\alpha \nu}  \, .
\ee
From this we infer the flow equation
\be
\det\[ \frac{\sqrt{-\det{g}}}{\sqrt{-\det{\gamma}} }  g^{\mu\alpha}(x) \gamma_{\alpha \nu} \]= 1 = e^{-4 \chi} \det[\alpha  \delta^{\mu}_{\nu} -  2 T^{\mu}{}_{\nu}] \, ,
\ee
which is to say $e^{4 \chi} = \alpha^2 - 2\alpha \Theta + 4\det T ^{\mu}{}_{\nu} $. Now
\ba
\frac{\d}{\d \lambda} S_{\bar T \bar T} &=& \int \d^2 x \sqrt{-\det \gamma} \[ \frac{1}{2} \frac{\partial \alpha}{\partial \lambda}  + \frac{\partial \chi}{\partial \lambda} \( \Theta -  \alpha\) \] \nn \\
&=& \int \d^2 x \sqrt{-\det \gamma} \[ \frac{1}{2} \frac{\partial \alpha}{\partial \lambda}  +\frac{1}{2\alpha} \frac{\partial \chi}{\partial \lambda} \(-\alpha^2 + 4 \det T ^{\mu}{}_{\nu} - e^{4 \chi}\) \] \, .
\ea
In order to interpret this purely as a $T \bar T$ deformation we need to have
\ba
 \frac{\partial \alpha}{\partial \lambda}  -\frac{1}{\alpha} \frac{\partial \chi}{\partial \lambda} \(\alpha^2 + e^{4 \chi}\)=0 \, , \quad  \frac{1}{\alpha} \frac{\partial \chi}{\partial \lambda}=-\frac{1}{2} \, ,
\ea
which gives the second order differential equation
\be\label{Specialsol}
\frac{\partial^2 \chi}{\partial \lambda^2} - \( \frac{\partial \chi}{\partial \lambda} \)^2 - \frac{1}{4} e^{4 \chi}=0 \, .
\ee
One solution of this equation is 
\be
e^{2\chi(x)}=\frac{1}{\lambda (1- \lambda V[\varphi])} \, , \quad \alpha(x) = \frac{1-2  \lambda V[\varphi]}{\lambda( 1- \lambda V[\varphi])} \, ,
\ee
which is precisely the solution considered in section \ref{specialconformal} given $\lambda = 1/m^2$. The most general solution is obtained from \eqref{Specialsol} by the replacement $\lambda \rightarrow \lambda + U(\varphi)$. It is easy to imagine how we might generalize arguments of this nature. \\

A crucial difference in the present case, distinct from that of a seed CFT, is that since the Polyakov and $ T\bar T$ amplitude are related by a $\varphi$ dependent cosmological constant term \eqref{TTPrelationspecial}, i.e. they are no longer equivalent theories even classically. The  $ T\bar T$ is equivalent to the non-critical string theory with an additional potential energy $\alpha( \varphi,\lambda)$ coupled to the target space metric. 
\ba
S_{\bar T \bar T} &=& \int \d^2 x  \, \[  - \frac{1}{4} \sqrt{-\det g}  g^{\mu\nu}  e^{2 \chi(\varphi,\lambda)}\hat \gamma_{AB}(\Phi) \partial_{\mu} \Phi^A \partial_{\nu} \Phi^B \]+ \nn \\
&&-\int \d^2 x  \, \frac{\partial \chi}{\partial \lambda}\sqrt{-\det \hat \gamma_{AB}(\Phi) \partial_{\mu} \Phi^A \partial_{\nu} \Phi^B }+S_{CFT}[\varphi,e] \, .
\ea
However, once again following the result of section \ref{Bfield}, this may be rewritten as
\be
S_{T\bar T}[\varphi,\gamma,e] =\int \d^2 x  \,  - \frac{m^2}{4} \left[ \sqrt{-g}  g^{\mu\nu} \partial_{\mu} \Phi^A \partial_{\nu} \Phi^B \hat \gamma'_{AB}( \varphi,\Phi) + B'_{AB}(\varphi,\Phi) \epsilon^{\mu\nu} \partial_{\mu} \Phi^A \partial_{\nu} \Phi^B \right] + S_{CFT}[\varphi,e] \, .
\ee
where
\be
\hat \gamma'_{AB}(\varphi,\Phi) =\lambda e^{2 \chi(\varphi,\lambda)} \hat \gamma_{AB}(\Phi)\, \quad \text{and  }  \quad B'_{AB}(\varphi,\Phi) =-2 \lambda  \frac{\partial \chi}{\partial \lambda}B_{AB}(\Phi)  \, .
\ee
with $B_{AB}(\Phi)  $ specified by $B_{+-}(\Phi)=\hat\gamma_{+-}(\Phi)$ in conformally flat gauge \eqref{conformallyflat}. This reproduces the results of section \ref{specialconformal}.

\section{Stochastic Path Integral for $J\bar T$ +$T \bar J$ + $T \bar T$ Deformations.}\label{generalstochastic}

In addition to the $T \bar T$ deformations, it is possible to consider deformations defined by any pair of conserved currents $J$ and $J'$ \cite{Cardy:2018sdv}. The simplest such examples are the $J \bar T$ and $T \bar J$ deformations first considered in \cite{Guica:2017lia}, but in turn it is possible to consider higher spin deformations \cite{Smirnov:2016lqw,Conti:2019dxg,LeFloch:2019wlf}. We will focus here on the case of the $J \bar T$ and $T \bar J$ deformations for simplicity. Explicit worldsheet strong constructions have been given in for example \cite{Chakraborty:2019mdf}. It is natural to ask if there is a path integral expression which describes the deformations in these cases. It is in fact straightforward to derive one following the method discussed in section \ref{Randomgeometry}. \\

To set notation, consider a pair of conserved currents $J_{\mu}$ and $\tilde J_{\mu}$, and define a pair of null vectors $n_{a}$ and $\tilde n_a$ which satisfy $n^2=\tilde n^2 =0$, $\epsilon^{ab} n_a n_b=1$ and $n.\tilde n=1$. To keep things relatively general, consider a combined $T \bar T$, $J \bar T$ and $ T \bar J $ deformation with relative weights $\alpha_{TT}$, $\alpha_{JT}$ and $\tilde \alpha_{JT}$, each of which are constants. 
The currents $J_{\mu}$ and $\tilde J_{\mu}$ are associated to some $U(1)$ global symmetry. In order to describe this in a path integral we gauge the symmetries by introducing background $U(1)$ gauge fields $A_{\mu}$ and $\tilde A_{\mu}$ which are the $U(1)$ version of $f_{\mu}^a$. At the level of the path integral $Z_{\lambda}[f,A,\tilde A,J_{\varphi}]$, where $J_{\varphi}$ describes external sources for $\varphi$,  the deformation is defined by the solution of the functional differential equation
\ba\label{functionalequation2}
i \frac{\d Z_{\lambda}[f,A,\tilde A,J_{\varphi}]}{\d \lambda} &=& (-i)^2 \int \d^2 x \lim_{y \rightarrow x}  \,  \left[   \frac{\alpha_{TT}}{2}  \epsilon_{\mu\nu} \epsilon^{ab} \frac{\delta^2 Z_{\lambda}[f,A,\tilde A,J_{\varphi}]}{\delta f_{\mu}^a(x) \delta f_{\nu}^b(y)} \right.  \nn \\
& +& \left. \alpha_{JT}  \epsilon_{\mu\nu} n^a \frac{\delta^2 Z_{\lambda}[f,A,\tilde A,J_{\varphi}]}{\delta A_{\mu}(x) \delta f_{\nu}^a(y) }+ \tilde \alpha_{JT} \epsilon_{\mu\nu} \tilde n^a \frac{\delta^2 Z_{\lambda}[f,A,\tilde A,J_{\varphi}]}{\delta \tilde A_{\mu}(x) \delta f_{\nu}^a(y) } \right]\, ,
\ea
corresponding to at the classical level a deformation
\be
\frac{\d S_{\lambda}[f,A,\tilde A ]}{\d \lambda} = - \int \d ^2 x \epsilon^{\mu \nu} \( \frac{1}{2} \alpha_{T \bar T} \epsilon_{ab} T_{\mu}^a T_{\nu}^b + \alpha_{JT}  n_a J_{\mu} T_{\nu}^a + \tilde \alpha_{JT}  \tilde n_a \tilde J_{\mu} T_{\nu}^a  \)  \, ,
\ee
given the classical identification
\be
 \det f  J^{\mu} =\frac{\delta S}{\delta A_{\mu}(x)} \,,   \quad \det f  \tilde J^{\mu} =\frac{\delta S}{\delta \tilde A_{\mu}(x)}  \, ,  \quad \det f T^{\mu}_a =\frac{\delta S}{\delta f^a_{\mu}(x)} \, .
\ee
Once again, \eqref{functionalequation2} is a stochastic, or functional Schr\"odinger equation in one additional dimension $\lambda$, with a Hamiltonian that is purely quadratic in momenta. We may hence immediately write down its path integral solution
\ba
&& Z_{\lambda_f}[f(\lambda_f),A(\lambda_f),\tilde A(\lambda_f),J_{\varphi}] = \int D f[x,\lambda] \int D P[x,\lambda] \int D P_A[x,\lambda]  \int D A[x,\lambda] \int D \tilde P_A[x,\lambda]  \int D \tilde A[x,\lambda] \nn \\
&& \exp\[ i \int_0^{\lambda_f} \d \lambda  \int \d^2  x  \[   P^{\mu}_a \partial_{\lambda} f_{\mu}^a+ P_A^{\mu} \partial_{\lambda} A_{\mu}+ {\tilde P_{ A}}^{\mu} \partial_{\lambda} \tilde A_{\mu} - \hat{\cal H} \] \] Z_{0}[f(0),A(0),\tilde A(0),J_{\varphi}]  \, , 
\ea
where 
\be
\hat {\cal H}= \frac{\alpha_{TT}}{2} \epsilon_{\mu\nu} \epsilon^{ab} P^{\mu}_a P^{\nu}_b+  \alpha_{JT}  \epsilon_{\mu\nu} n^a P_A^{\mu}  P^{\nu}_a+\tilde \alpha_{JT}\epsilon_{\mu\nu} \tilde n^a \tilde P_A^{\mu}  P^{\nu}_a \, .
\ee
Performing the path integral over $f$, $A$ and $\tilde A$ which each give functional delta functions, then remembering to keep track of boundary terms, and performing an obvious change of notation we have
\ba
 &&  \hspace{-15pt} Z_{\lambda}[f,A,\tilde A,J_{\varphi}] = \int D e[x] \int D P[x] \int D P_A[x]  \int D B[x]  \int D \tilde P_A[x] \int D \tilde B[x]  \nn \\
&&  \hspace{-15pt}  \exp\[i    \int \d^2 x  \(  P^{\mu}_a (f_{\mu}^a-e_{\mu}^a)+   P_A^{\mu} (A_{\mu}-B_{\mu}) +   {\tilde P_A}^{\mu} (\tilde A_{\mu}-\tilde B_{\mu})  -\tilde {\cal H}     \)\]Z_{0}[e,B,\tilde B,J_{\varphi}] \, ,
\ea
with
\be
\tilde {\cal H} = \lambda \[ \frac{\alpha_{TT}}{2} \epsilon_{\mu\nu} \epsilon^{ab} P^{\mu}_a P^{\nu}_b+  \alpha_{JT}  \epsilon_{\mu\nu} n^a P_A^{\mu}  P^{\nu}_a+\tilde \alpha_{JT}\epsilon_{\mu\nu} \tilde n^a \tilde P_A^{\mu}  P^{\nu}_a \] \, ,
\ee
with the difference being that now $\tilde {\cal H}$ is a function of momentum fields which are just functions of $x$, and the path integrals are similarly only over functions of $x$. \\

Stated differently, the classical theory which we need to quantize to describe a combined $T\bar T+J\bar T+T \bar J$ deformation is
\be
S_{T\bar TJ\bar T J\bar T}= \int \d^2 x  \(  P^{\mu}_a (f_{\mu}^a-e_{\mu}^a)+   {P_A}^{\mu} (A_{\mu}-B_{\mu}) +   {\tilde P_A}^{\mu} (\tilde A_{\mu}-\tilde B_{\mu})  -\tilde {\cal H} \)  +S_0[e,B,\tilde B] \, .
\ee
How we proceed depends on the precise structure of the deformations, i.e. it is slightly different if certain coefficients vanish. For simplicity, we consider the most general case for which all the deformations are non-zero. 
Integrating over ${P_A}^{\mu}$ and ${\tilde P_A}^{\mu}$ sets two constraints
\be
A_{\mu}- B_{\mu}- \lambda \alpha_{JT}  \epsilon_{\mu\nu} n^a P^{\nu}_a=0 \, , \quad  \tilde A_{\mu}- \tilde B_{\mu}- \lambda \tilde \alpha_{JT}  \epsilon_{\mu\nu} \tilde n^a P^{\nu}_a  =0 \, .
\ee
Provided $\alpha_{JT} \neq 0$ and $\tilde \alpha_{JT} \neq 0$ we may view these as fixing $P_{\mu}^a$, namely
\be
P^{\mu}_a =- \frac{1}{\lambda \alpha_{JT} } \epsilon^{\mu\nu} \tilde n_a (A_{\nu}- B_{\nu}) -\frac{1}{\lambda \tilde \alpha_{JT} } \epsilon^{\mu\nu} n_a (\tilde A_{\nu}- \tilde B_{\nu})  \, .
\ee
This removes the final momentum integrals, giving
\ba
 &&  \hspace{-15pt} Z_{\lambda}[f,A,\tilde A,J_{\varphi}] = \int D e[x] \int D B[x]   \int D \tilde B[x]  \exp\[i    \int \d^2 x  \(-  \frac{1}{\lambda \alpha_{JT} } \epsilon^{\mu\nu} \tilde n_a (f_{\mu}^a-e_{\mu}^a)(A_{\nu}-B_{\nu})   \right.  \right.  \nn \\
&& \hspace{-15pt}  \left. \left. -  \frac{1}{\lambda \tilde \alpha_{JT} } \epsilon^{\mu\nu} n_a  (f_{\mu}^a-e_{\mu}^a)(\tilde A_{\nu}-\tilde B_{\nu}) -   \frac{\alpha_{TT}}{ \lambda \alpha_{JT} \tilde \alpha_{JT}} \epsilon^{\mu\nu} (A_{\mu}-B_{\mu})(\tilde A_{\nu}-\tilde B_{\nu})    \)\]Z_{0}[e,B,\tilde B,J_{\varphi}] \, . 
\ea
This is exactly the path integral considered recently in \cite{Aguilera-Damia:2019tpe}, at least in the case where the reference metric is Minkowski. More precisely  \cite{Aguilera-Damia:2019tpe} consider a St\"uckelberg-erized form in which we introduce a $U(1)$ \stu field in the $B$ and $\tilde B$
\be
B_{\mu} \rightarrow B_{\mu} - \partial_{\mu} \alpha \, ,  \quad \tilde B_{\mu} \rightarrow \tilde B_{\mu} - \partial_{\mu} \tilde \alpha  \, ,
\ee
and linear diffeomorphism \stu fields $e_{\mu}^a \rightarrow e_{\mu}^a - \partial_{\mu} Y^a$. The path integral will then include the integral over the \stu fields and a division by the volume of the gauge orbit. Although it makes no difference to the unitary gauge Lagrangian, from our perspective it makes more sense to introduce standard non-linear diffeomorphism \stu fields through $f^a_{\mu} \rightarrow F^a_{A}(\Phi) \partial_{\mu} \Phi^A$, in particular to account for the case in which the spacetime is curved. However we suspect that for at least the flat reference metric case, this will make no difference to the essential results of  \cite{Aguilera-Damia:2019tpe} given that these symmetries are broken by the deformation terms, and we are free to introduce \stu fields however we choose without fundamentally change the physics. The only difference between them is the precise way in which we deal with the volume of the gauge orbit. We refer to \cite{Aguilera-Damia:2019tpe}, and \cite{Anous:2019osb} in the special case of a pure $J\bar T$ deformation, for a more complete treatment of the quantum path integral.  It is straightforward to see how these arguments may be generalized to the higher spin deformations  \cite{Smirnov:2016lqw,Conti:2019dxg,LeFloch:2019wlf}.

\section{Discussion}

In this article we have shown that classically the $T \bar T$ deformation of an arbitrary field theory on curved spacetime is equivalent to two dimensional ghost-free massive gravity, and that when the field theory is a CFT this in turn is classically equivalent to a non-critical string theory. In the process we have been able to rederive many relations previously observed in the literature, but with considerable ease. For instance, we are able to give a closed form to the action for the classical $T \bar T$ deformation of an arbitrary CFT. We are also able to derive the deformation straightforwardly as a consequence of the action for a non-critical string, without making use of gauge fixing, which has been the approach previously relied on. \\

We have also given explicit solutions to proposed versions of the quantum flow equations in terms of a stochastic path integral in which the flow parameter $\lambda$ acts as an additional spacetime direction. This result was implicit in the work of \cite{Cardy:2018sdv}, however our derivation is considerably more transparent, and resolves issues about the choice of measure. In particular we have seen that we reproduce exactly the path integrals proposed for $T \bar T$ deformations in \cite{Dubovsky:2017cnj,Dubovsky:2018bmo} and for more general deformations in \cite{Aguilera-Damia:2019tpe} with considerable ease, and this approach can straightforwardly be generalized to arbitrary higher spin deformations. \\

The central question is -- to what extent does our proposal survive quantization on a curved spacetime? When the reference metric is flat, the proposed path integrals have passed several non-trivial checks, but on a curved spacetime the conformal anomaly, i.e. the central charge,  becomes important, and so many of the formal manipulations may lead to non-trivial measure contributions. We may however take the perspective that we define the quantum theory by whatever classical description appears to be simplest to quantize. In that vain, there is a clear proposal for the case where the seed theory is a CFT, namely as the quantized description of a non-critical string along the lines of the flat space proposal considered recently in \cite{Callebaut:2019omt}. That is because, considering the original CFT on curved spacetime is equivalent to the very modest deformation of a non-critical string with curved target space metric. The latter, although not without issues, is much better understood and there has been notable work on non-critical string quantization \cite{Polchinski:1991ax,Dubovsky:2012sh,Hellerman:2014cba,Dodelson:2017emn,Callebaut:2019omt}. We leave it to future work to test these proposals.

\bigskip
\noindent{\textbf{Acknowledgments:}}

We would like to thank Lasma Alberte, Claudia de Rham, Laurent Freidel, Yunfeng Jiang, Vasudev Shyam, Arkady Tseytlin and Toby Wiseman for useful comments. The work of AJT is supported by an STFC grant ST/P000762/1. AJT thanks the Royal Society for support at ICL through a Wolfson Research Merit Award.

\bigskip

\appendix

\section{Path Integral Measure}\label{app:measure}

The measure of the path integral can be understood as the volume measure associated with a metric on superspace, in this case the space of all possible zweibein configurations\footnote{We would like to thank Vasudev Shyam for helpful discussions that prompted this analysis.}. The natural metric over zweibeins is
\be
\delta s^2 =- \int \d^2 x \,  \epsilon^{\mu \nu} \epsilon_{ab} \delta e_{\mu}^a(x)  \delta e_{\nu}^b(x)=  -\int \d^2 x \, 2 \det[ \delta e_{\mu}^a(x)]\, . 
\ee
This superspace metric is diffeomorphism invariant by virtue of being the integral of a two-form. The determinant of this superspace metric is essentially unity (or a fixed constant) since for each $x$
\be
\det[\epsilon \otimes \epsilon] =1 \, ,
\ee
and the full determinant on superspace is
\be
{\rm Det}[G^{\mu}{}_{a x ;}{}^{\nu}{}_{b y} ] = \Pi_x \det[\epsilon \otimes \epsilon] =\Pi_x  1 =1\, .
\ee
Since the determinant is a constant, the path integral over zweibeins preserves the linearity property
\be
\int \D e = \int \D (e+f) \, .
\ee
Furthermore up to an overall normalization the Gaussian integral is defined as
\be
\int \D e  \, e^{ \frac{i}{2} \int \d^2 x \,  \epsilon^{\mu \nu} \epsilon_{ab} (e_{\mu}^a(x) - f_{\mu}^a(x))(  e_{\nu}^b(x)- f_{\nu}^b(x))} = 1 \, .
\ee
This is what is implicitly used in sections \ref{Randomgeometry} and \ref{generalstochastic}. It is not immediately clear that this linear measure for zweibeins is equivalent to the standard measure for integration over metrics used for example in quantizing a bosonic string \cite{Polyakov:1981rd}. Fortunately it is, as can be seen by changing variables in superspace.  A generic zweibein can be thought of as a metric perturbation together with a local Lorentz transformation. At the infinitessimal level, this implies the following decomposition
\be
\delta e_{\mu}^a = \eta^{ac} \epsilon_{cd} \delta  \omega \, e_{\mu}^d + \delta h_{\mu}^a \, ,
\ee
where $\delta \omega$ denotes the infinitessimal form of a local Lorentz transformation $\Lambda^a{}_b = e^{\eta^{ac} \epsilon_{cb} \delta \omega}$, and $\delta h_{\mu}^a$ encodes the remaining metric perturbations. In order to directly associate $\delta h_{\mu}^a$ with the metric perturbation we must fix a local Lorentz gauge, and the best one to choose is the symmetric vielbein condition
\be
\eta_{ab} e_{\mu}^a \delta h_{\nu}^b = \eta_{ab} e_{\nu}^a \delta h_{\mu}^b \, .
\ee
Crucially with this choice we have the factorization
\be
\det[\eta^{ac} \epsilon_{cd} \delta \omega \, e_{\mu}^d + \delta h_{\mu}^a]= \det[\eta^{ac} \epsilon_{cd} \delta \omega \, e_{\mu}^d] + \det[\delta h_{\mu}^a] = - \delta \omega^2 \det(e)+\det[\delta h_{\mu}^a]  \, .
\ee
Our goal now is to rewrite this in terms of the actual metric perturbation $\delta g_{\mu \nu}$. Since $g_{\mu\nu} = \eta_{ab}e_{\mu}^a e_{\nu}^b$ then
\be
\delta g_{\mu\nu} = \eta_{ab} e_{\mu}^a \delta h_{\nu}^b +   \eta_{ab} \delta h_{\mu}^a e_{\nu}^b  \, ,
\ee
where the local Lorentz transformations $\delta \omega$ drop out. By virtue of the symmetric vielbein condition this is
\be
\delta g_{\mu\nu}  = 2  \eta_{ab} e_{\mu}^a \delta h_{\nu}^b \, .
\ee
From this we infer 
\be
g^{\mu \alpha} \delta g_{\alpha \nu} = 2 e^{\mu}_a \delta h_{\nu}^a \, ,
\ee
and so on taking the determinant
\be
\det[g^{\mu \alpha} \delta g_{\alpha \nu}] = 4 \det[e^{\mu}_a \delta h_{\nu}^a] = \frac{4}{\det(e)} \det[\delta h_{\mu}^a] \, .
\ee
Putting this together (remembering that $\det (e) =   \sqrt{-g}$) we have
\ba\label{finalsuperspace}
\delta s^2 &=& \int \d^2 x  \sqrt{-g} \[  2  \delta \omega^2  - \frac{1}{2} \det[g^{\mu \alpha} \delta g_{\alpha \nu}]  \]  \, ,\\
&=& \int \d^2 x  \sqrt{-g} \[ 2  \delta \omega^2  +\frac{1}{4}  \( g^{\mu \nu} g^{\alpha \beta}\delta g_{\mu \alpha} g_{\nu \beta} \) - \frac{1}{4} \( g^{\mu \nu} \delta g_{\mu \nu} \)^2  \]  \, .
\ea
The first term is the naive covariant measure for integration over local Lorentz transformations in two dimensions. The second and third term are exactly Polyakov's measure for the bosonic string with the special choice $C=-1$ (in the notation of  \cite{Polyakov:1981rd}). Most importantly this measure is manifestly diffeomorphism invariant and includes within it the Liouville measure.
Indeed borrowing the argument of \cite{Polyakov:1981rd}, the `unitary' gauge path integral measure then becomes
\be\label{finalmeasure}
\int \D e(x)  \equiv \int \D \omega(x) \int \D \sigma(x) \int D \xi^a(x) \, \sqrt{\det \hat {\cal L}} \, ,
\ee
where $e^{\sigma(x)}$ is the conformal factor of the two dimensional metric, $\xi^a(x)$ encodes diffeomorphisms and $\sqrt{\det \hat{\cal L}}$ gives the Liouville measure where
\be
\( \hat{\cal L} \xi\)^a= \nabla_b (\nabla^a \xi^b+ \nabla^b \xi^a - g^{ab} \nabla_c \xi^c) \, .
\ee
All the integrals on the right hand side of \eqref{finalmeasure} are defined in a manifestly diffeomorphism invariant way (i.e. including factors of $\sqrt{-g}$ as per \eqref{finalsuperspace}). The only difference from  \cite{Polyakov:1981rd} is the extra integration over local Lorentz transformations. Equation \eqref{finalmeasure} is the unitary gauge version of the appropriate path integral measure that comes from the \stu formulation in which $\int \D \omega(x) $ describes the integration over the local Lorentz \stu fields $\Lambda^a{}_b(x)$, $ \int D \xi^a(x)$ the integration over the local diffeomorphism \stu fields $\Phi^a$, and the two dimensional metric is gauge fixed to be conformally flat $g_{\mu\nu} = e^\sigma(x) \eta_{\mu\nu}$. It is somewhat remarkable that the nontrivial Liouville measure is hidden within the rather standard linear measure over zweibeins.

\bibliographystyle{JHEP}
\bibliography{references}

\end{document}